\documentclass[preprint,review,12pt]{elsarticle}

\usepackage{amssymb}
\usepackage{subfigure}
\usepackage{xcolor}
\usepackage{booktabs}
\usepackage{hyperref}
\usepackage{amsmath}
\usepackage{booktabs}
\usepackage{ulem}
\usepackage{siunitx}

\usepackage{lineno}

\journal{Nuclear Instruments and Methods in Physics Research A}

\begin{document}

\begin{frontmatter}

\title{Construction and characterization of a muon trigger detector for the PSI muEDM experiment}

\author[sjtu]{Guan~Ming~Wong}
\author[sjtu]{Tianqi~Hu}
\author[sjtu]{Samip~Basnet\fnref{sbnote}}
\author[psi]{Chavdar Dutsov}
\author[sjtu]{Siew Yan Hoh\fnref{syhnote}}
\author[psi,ethz]{David H\"ohl}
\author[sjtu]{Xingyun Huang}
\author[psi,ethz]{Timothy David Hume}
\author[psi]{Alexander Johannes J\"ager}
\author[sjtu]{Kim~Siang~Khaw\corref{ksknote}}
\author[sjtu]{Meng Lyu\fnref{lmnote}}
\author[psi]{Ljiljana Morvaj}
\author[sjtu]{Jun~Kai~Ng}
\author[pisa,infn,psi]{Angela Papa}
\author[psi]{Diego Alejandro Sanz Becerra}
\author[psi]{Philipp Schmidt-Wellenburg}
\author[sjtu]{Yusuke~Takeuchi}
\author[sjtu]{Yonghao~Zeng}

\cortext[ksknote]{Corresponding author: kimsiang84@sjtu.edu.cn}
\fntext[sbnote]{Present address: Institute of Experimental and Applied Physics, Czech Technical University in Prague, Husova 5, Prague, Czech Republic}
\fntext[syhnote]{Present address: Department of Physics, Xiamen University Malaysia, 43900 Sepang, Selangor, Malaysia.}
\fntext[lmnote]{Present address: Graduate School of Science, The University of Tokyo, 7-3-1 Hongo, Bunkyo-ku, Tokyo 113-0033, Japan}
\address[sjtu]{State Key Laboratory of Dark Matter Physics, Key Laboratory for Particle Astrophysics and Cosmology (MOE), Shanghai Key Laboratory for Particle Physics and Cosmology (SKLPPC),  Tsung-Dao Lee Institute \& School of Physics and Astronomy, Shanghai Jiao Tong University, Shanghai 201210, China}
\address[psi]{PSI Center for Neutron and Muon Sciences, 5232 Villigen PSI, Switzerland}
\address[ethz]{ETH Zürich, 8093, Zurich, Switzerland}
\address[pisa]{Dipartimento di Fisica, Università di Pisa, Largo B. Pontecorvo 3, Pisa, 56127, Italy}
\address[infn]{Sezione di Pisa, Instituto Nazionale di Fisica Nucleare, Largo B. Pontecorvo 3, Pisa, 56127, Italy}

\begin{abstract}
We present the upgraded design, construction, and beam test results for the Muon Trigger Detector (MTD) developed for the muon Electric Dipole Moment (muEDM) experiment at the Paul Scherrer Institute (PSI) in Switzerland. This experiment aims to improve the sensitivity of the muon EDM measurement by more than three orders of magnitude beyond the current limit established by the BNL Muon $g-2$ experiment. Precise identification of storable incoming muons at the entrance of the storage solenoid is essential, as the MTD must rapidly trigger a pulsed magnetic kicker to confine muons in the central region of the solenoid, where a weakly focusing magnetic field is maintained. The MTD comprises two subsystems: a \SI{0.1}{mm}-thick plastic scintillator ``gate detector'' read out by four silicon photomultipliers (SiPMs), and a \SI{5}{mm}-thick CNC-machined plastic scintillator ``active aperture detector'' read out by six SiPMs. The geometry of the active aperture detector was optimized through acceptance studies to maximize both storage efficiency and background veto efficiency. Integrated fast electronics generate an LVTTL trigger signal under an anti-coincidence condition---a muon detected in the gate but not in the aperture---ensuring selective triggering of storable muon events for the EDM measurement. The system was tested at the PSI $\pi$E1 beamline using \SI{22.5}{MeV/\textit{c}} muons under scaled-down conditions to characterize detector response and trigger performance. A Geant4 simulation incorporating detailed optical photon transport and SiPM response modeling was developed and reproduces the measured event topologies with ${\sim}97\%$ agreement. These results validate the detector design and demonstrate the MTD's readiness for deployment in the full muEDM Phase-1 setup.
\end{abstract}



\end{frontmatter}


\section{Introduction}
\label{sec:introduction}

The search for the muon electric dipole moment (EDM) constitutes a powerful probe of charge-parity (CP) symmetry violation beyond the Standard Model (SM). The muon EDM search complements searches for the electron, neutron, and mercury EDMs, each probing different aspects of CP violation, with the muon offering a clean probe of a bare second-generation lepton, free from atomic or nuclear modeling uncertainties~\cite{Chupp:2017bad, Roussy:2023jig, Abel:2020gbr, Graner:2016vnx}. The most stringent experimental upper limit to date, $d_\mu < 1.8 \times 10^{-19}~e\cdot$cm (95\% CL), was established by the BNL E821 experiment~\cite{Muong-2:2008ebm}---a value 19 orders of magnitude above theoretical SM expectations~\cite{Pospelov:2013sca, Ghosh:2017uqq, Yamaguchi:2020eub}. The PSI muEDM experiment~\cite{Adelmann:2025nev} employs the frozen-spin technique~\cite{Farley:2003wt}, in which the anomalous magnetic moment's contribution to spin precession is cancelled by a carefully configured electric field within a storage solenoid. This approach targets an unprecedented muon EDM sensitivity of $6 \times 10^{-23}~e\cdot$cm, offering a compelling opportunity to discover signatures of new physics, including those arising from undiscovered particles or interactions~\cite{Khaw:2022qxh,Nakai:2022vgp,Crivellin:2018qmi}.

\begin{figure}[htbp]
    \centering
    \includegraphics[width=0.8\linewidth]{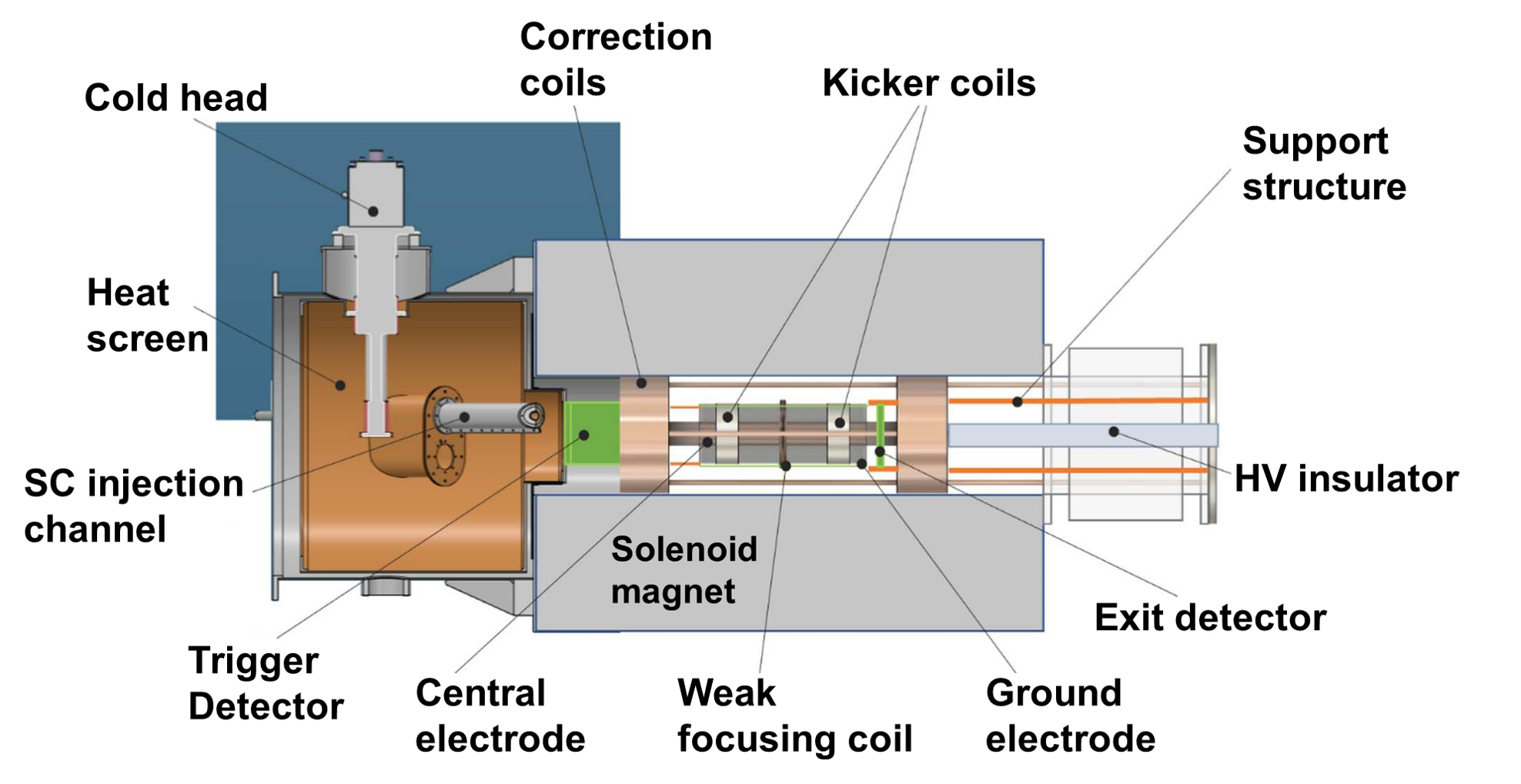}
    \caption{Layout of muEDM Phase-1 experimental setup, showing side-view.}
    \label{fig:muedmPhase1}
\end{figure}

The muEDM experiment at PSI proceeds in two phases, with Phase-1 serving as a demonstrator of the frozen-spin technique incorporating all key experimental components, as depicted in Fig.~\ref{fig:muedmPhase1}. While Phase-1 uses a surface muon beam and a storage solenoid at \SI{3}{T}, Phase~2 will employ a dedicated instrument designed for two orders of magnitude higher sensitivity by better matching the trap's storage phase space to the incident muon phase space at a momentum of up to \SI{140}{MeV/c}. Phase-1 utilizes a surface muon beam from the $\pi$E1 beamline and a storage solenoid operating at approximately \SI{3}{T}. Muons are injected off-axis~\cite{Iinuma:2016zfu} through a superconducting channel~\cite{muEDM:2023mtc} shielding them from the magnetic fringe field of the storage solenoid. Upon exiting the injection tubes, muons spiral through the Muon Trigger Detector (MTD), which generates a fast signal directed to a pulsed power supply in the central storage region. Upon receipt of this trigger signal, the pulsed power supply generates a current generating a magnetic pulse reducing the muons' longitudinal motion, capturing them on a stable storage orbit. Within this storage region, the frozen-spin electric field will be applied radially~\cite{muEDM:2024bri}, cancelling spin precession contributions from the muon's anomalous magnetic moment, thereby enabling a pure probe of the muon EDM. The EDM is subsequently extracted from the change of the longitudinal asymmetry with time.

The MTD plays a pivotal role in enabling the muon EDM measurement. In this work, we present the deployment of an upgraded MTD prototype, building upon a proven proof-of-concept design~\cite{Hu:2025tgk, Hu:2025egg, Stager:2023lgy} and developed for installation in the PSC solenoid with optimization for axial beam injection. This prototype incorporates a thin plastic scintillator serving as the Gate to detect incoming muons, combined with a thick plastic scintillator functioning as an Active Aperture to stop and veto out-of-acceptance muons. Both the Gate detector (GD) and Active Aperture detector (AAD) are read out using silicon photomultipliers (SiPMs) and integrated with a fast-electronics readout system that generates a Low-Voltage Transistor-Transistor Logic (LVTTL) trigger signal. A comprehensive beam test was conducted at the $\pi$E1 beamline at PSI in October 2024 to assess the MTD's performance under various experimental conditions.

This paper is organized as follows: Section~\ref{sec:MTD} introduces the Muon Trigger Detector and its experimental requirements; Section~\ref{sec:design} describes the detector design informed by Geant4-based simulations, the fabrication of the aperture, and in-laboratory characterization tests; Section~\ref{sec:setup} presents the experimental setup and measurements conducted during the PSI Test Beam 2024 campaign; Section~\ref{sec:results} discusses the detector's performance on the basis of measurement--simulation agreement; and Section~\ref{sec:conclusion} concludes with a summary of findings regarding this prototype's performance.

\section{Muon Trigger Detector}
\label{sec:MTD}

The MTD, depicted in Fig.~\ref{fig:MTDSketch}, is a fast and precise trigger system designed to activate the pulsed power supply and thereby ensure efficient muon storage in the muEDM experiment. It comprises a \SI{0.1}{mm}-thick Gate and a \SI{5}{mm}-thick Active Aperture, both fabricated from plastic scintillator and read out by SiPMs. A trigger signal is generated when a muon is detected by the Gate but not by the Active Aperture---an anti-coincidence condition---ensuring the selective identification of storable muons. This selectivity is critical: the incident muon rate following the superconducting tube is \SI{120}{kHz}, of which only 0.4\% are within the storage phase space. To satisfy the muEDM experimental requirements, the MTD must fulfill the following criteria:

\begin{figure}[htbp]
    \centering
    \includegraphics[width=\linewidth]{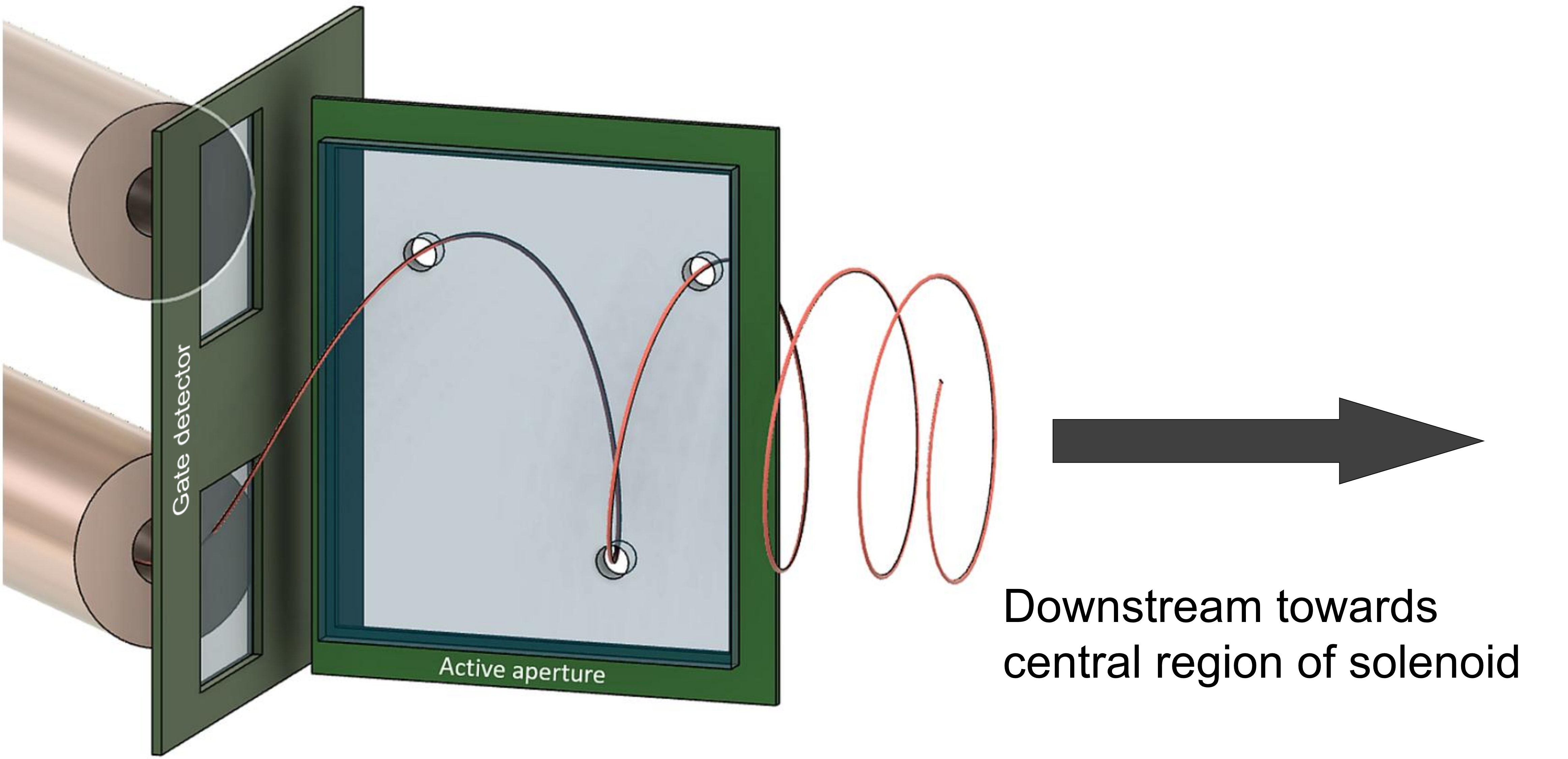}
    \caption{Conceptual sketch of the Muon Trigger Detector (MTD).}
    \label{fig:MTDSketch}
\end{figure}

\begin{enumerate}
    \item Fit within a compact cylindrical envelope with a diameter $\leq$\SI{200}{mm} and a length of $\leq$\SI{150}{mm};
    \item Introduce minimal perturbation to muon beam trajectories so as to maintain high storage efficiency;
    \item Achieve selective triggering with an acceptance rate $\varepsilon_a \geq 95\%$ and a rejection rate $\varepsilon_r \geq 98\%$;
    \item Provide a fast trigger signal to the pulsed power supply system, applying a magnetic pulse within $t_\mathrm{delay} \leq \SI{140}{ns}$ after muon passage.
\end{enumerate}

An earlier beam test of the proof-of-concept prototype~\cite{Hu:2025egg} demonstrated the feasibility of the anti-coincidence scheme and confirmed that plastic scintillators coupled with SiPMs constitute viable detector components, owing to their nanosecond-scale response times. The upgraded prototype aims to demonstrate fast, precise triggering under realistic conditions, while avoiding multiple scattering on additional detectors. The fast electronics component is described in a separate publication~\cite{Hu:2025tgk}; this study focuses on evaluating the precision requirements through measurements and simulations of $\varepsilon_a$ and $\varepsilon_r$.

The acceptance rate $\varepsilon_a$ directly influences the experimental sensitivity. 
A reduction in $\varepsilon_a$ correspondingly increases the data-taking time required to achieve the target sensitivity.


The rejection rate $\varepsilon_r$ is constrained by the pulsed power supply's rate limit, which is limited to a maximum repetition rate of $f_\mathrm{kick} \leq \SI{2}{kHz}$~\cite{Adelmann:2025nev}, while the injection rate is $f_\mathrm{inj} = \SI{120}{kHz}$ and the storable muon rate is $f_\mathrm{sto} = \SI{480}{Hz}$. This yields the constraint:
\begin{equation}
    \varepsilon_r \geq \frac{f_\mathrm{inj} - f_\mathrm{kick}}{f_\mathrm{inj} - f_\mathrm{sto}} \simeq 98\% \,.
    \label{eq:rejection}
\end{equation}
Without increasing the number of stored muons, even a 1\% shortfall in $\varepsilon_r$ would permit an additional \SI{1.2}{kHz} of non-storable muons to reach the pulsed power supply, exceeding its operational limit for triggered kicks.

While this work focuses on detector design and characterization, a comprehensive evaluation of how $\varepsilon_a$ and $\varepsilon_r$ impact EDM sensitivity and systematic uncertainties is essential and will be addressed in future dedicated studies.

\section{Design and construction of the muon trigger detector}
\label{sec:design}

\subsection{Detector design with simulations}
\label{sec:sim_design}

The baseline detector design was established using two Geant4-based~\cite{Allison:2016lfl} simulation frameworks: G4Beamline~\cite{Roberts:2007nte, g4bl} for beam transport and musrSim~\cite{musrsim} for detailed detector response. These frameworks operated in a complementary fashion, with G4Beamline handling beam dynamics and field propagation while musrSim modeled particle interactions within the detector volumes. 
The objective of the simulation study was to determine an aperture geometry that simultaneously maximizes the acceptance rate $\varepsilon_a$ and the rejection rate $\varepsilon_r$. In simulation, these metrics are evaluated from the detector response, specifically the energy deposition in the scintillators. To validate the simulation framework, the simulated photoelectron counts were benchmarked against the measured charge distributions after conversion to photoelectrons via calibration. Good agreement confirms that the simulation reliably models the physical detector response, thereby supporting the use of this simulation‑based design approach. While direct experimental determination of $\varepsilon_a$ and $\varepsilon_r$ was not part of the present test beam scope, which focused on detector response validation, the validated simulation framework provides confidence that these metrics are accurately modeled and will be addressed in future dedicated measurements under nominal conditions.

The full muEDM Phase-1 experimental setup was simulated to establish the baseline design. The simulation employed a set of injection parameters (Table~\ref{tab:injection}) optimized for muon storage efficiency $\varepsilon_s$, defined as the ratio of stored muons $N_s$ to injected muons $N_i$. Stored muons were defined as those that are stopped, decay, or are eliminated within the simulation volume at $z = \pm\SI{40}{mm}$. These optimal injection parameters, obtained via a surrogate model optimization algorithm based on polynomial chaos expansion (PCE), yielded a storage efficiency $\varepsilon_s$ of approximately 0.3\%.

\begin{table}[htbp]
\centering
\caption{Optimal injection parameters (injection parameters v1) obtained from the Polynomial Chaos Expansion (PCE) optimization procedure~\cite{Hoh:2025wws}.}
\label{tab:injection}
\begin{tabular}{@{}lc@{}}
\toprule
Parameter & Value \\
\midrule
Injection radius, $\mathrm{inj}_R$ (mm) & 45.561 \\
Injection vertex from solenoid centre, $m_Z$ (mm) & $-443.836$ \\
Polar angle, $\theta$ ($^\circ$) & $-45.022$ \\
Azimuthal angle, $\phi$ ($^\circ$) & 9.244 \\
Weak focusing current, WeakCurr (A/mm$^2$) & 1.5 \\
Correction coil current, SplitPair (A/mm$^2$) & 2.5 \\
\bottomrule
\end{tabular}
\end{table}

\begin{figure}
    \centering
    \includegraphics[width=\linewidth]{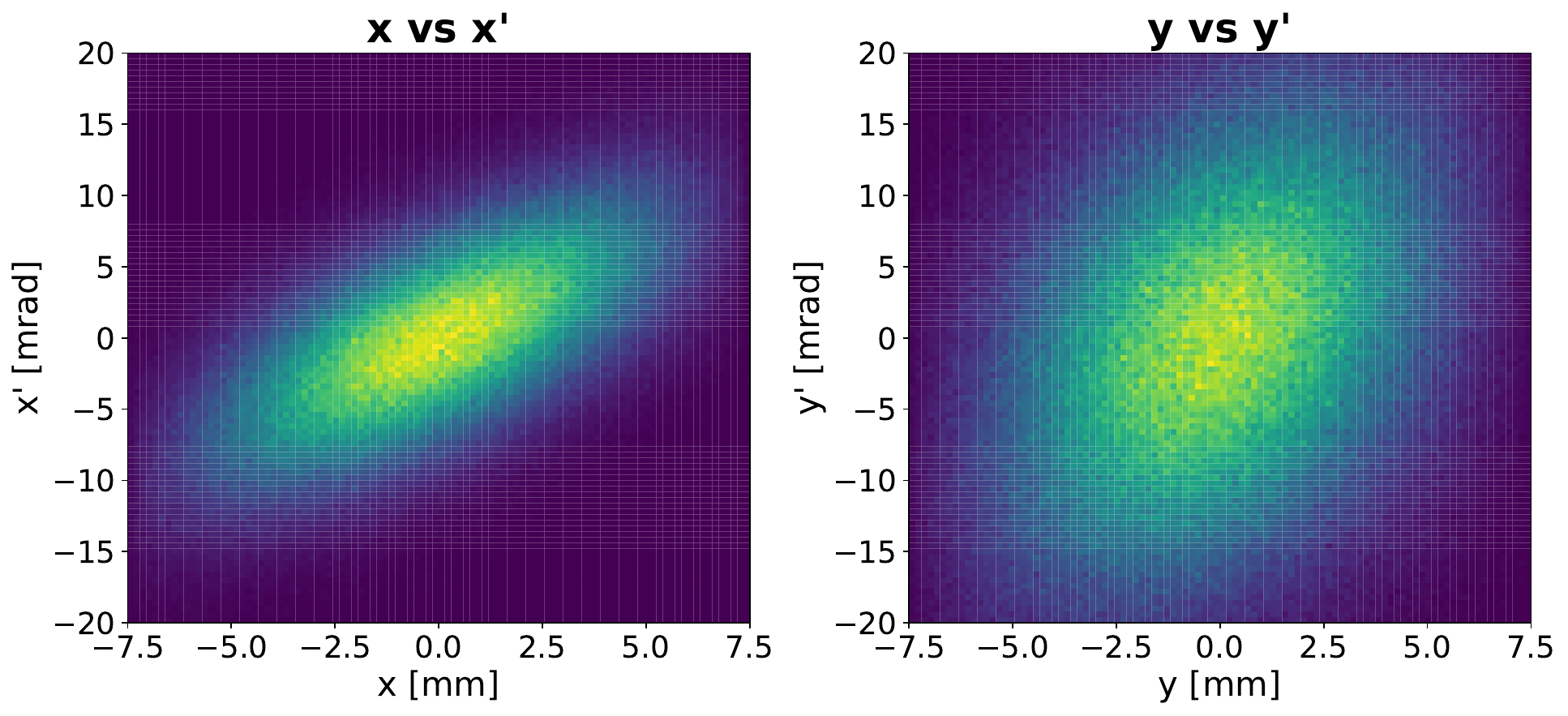}
    \caption{Monte-Carlo generated beam phase space at the exit of the injection tubes. The phase space distributions are based on a measurement at the $\pi$E1 beamline exit, propagated through a tube accounting only for geometric acceptance. Coordinates are given in the laboratory frame with origin at the center of the injection tube exit. Left: $x'$ vs $x$, Right: $y'$ vs $y$.}
    \label{fig:inj_dist}
\end{figure}

\begin{figure}
    \centering
    \includegraphics[width=\linewidth]{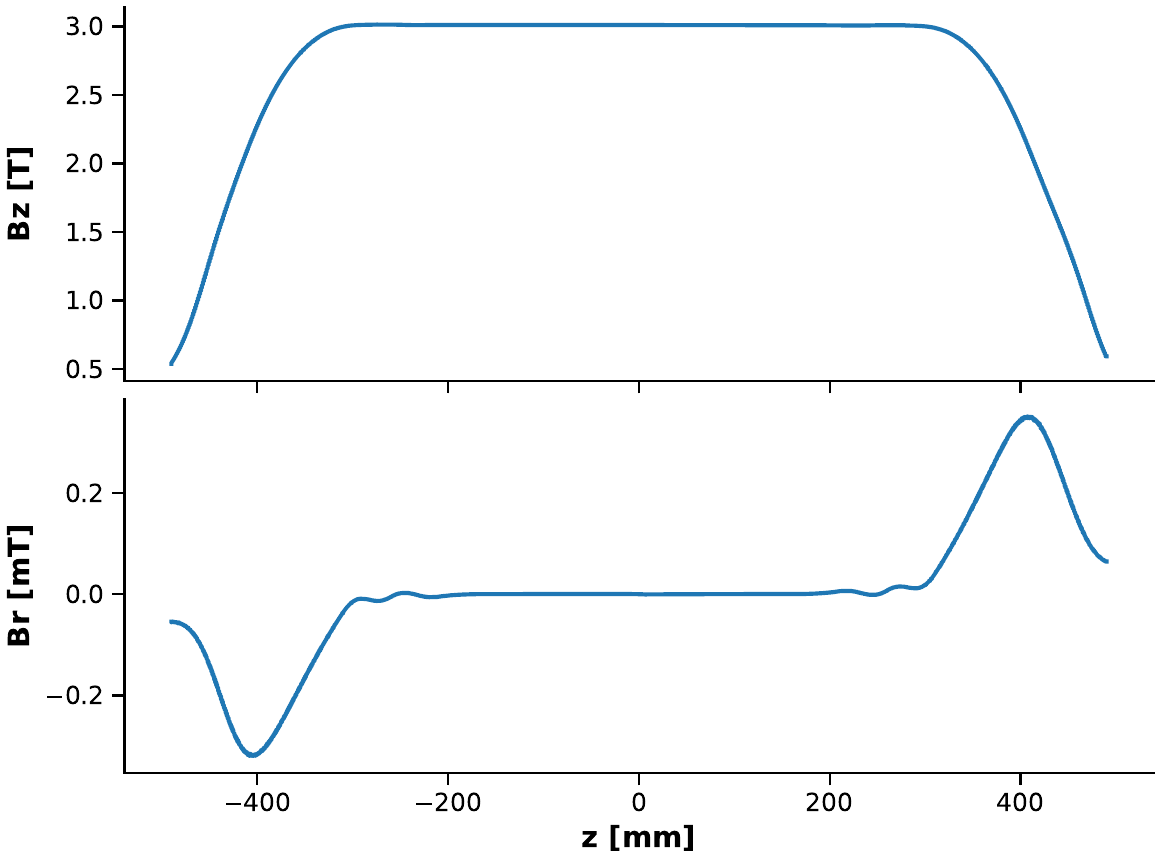}
    \caption{Magnetic fields implemented in simulation, comprising fields of the PSC solenoid, the Correction Coils, and the Weakly Focusing coil. Top: z-component along the solenoid axis; bottom: radial component. The coordinate origin is at the center of the solenoid.}
    \label{fig:fieldmap}
\end{figure}

Figure~\ref{fig:inj_dist} shows the beam phase space employed in the simulation. The setup used the surface muon beam from the $\pi$E1 beamline, characterized by its Twiss parameters. The beam was collimated through a tube of \SI{1}{m} length and \SI{15}{mm} diameter. All relevant magnetic fields (Fig.~\ref{fig:fieldmap}) were incorporated: the main \SI{3}{T} solenoid field, correction fields, the weakly focusing field, the pulsed power supply field, and the frozen-spin electric field. The main solenoid field was derived from simulation-interpolated measurement points, whereas the remaining fields were obtained from finite element method (FEM) simulations.

\begin{figure}[htbp]
    \centering
    \includegraphics[width=\linewidth]{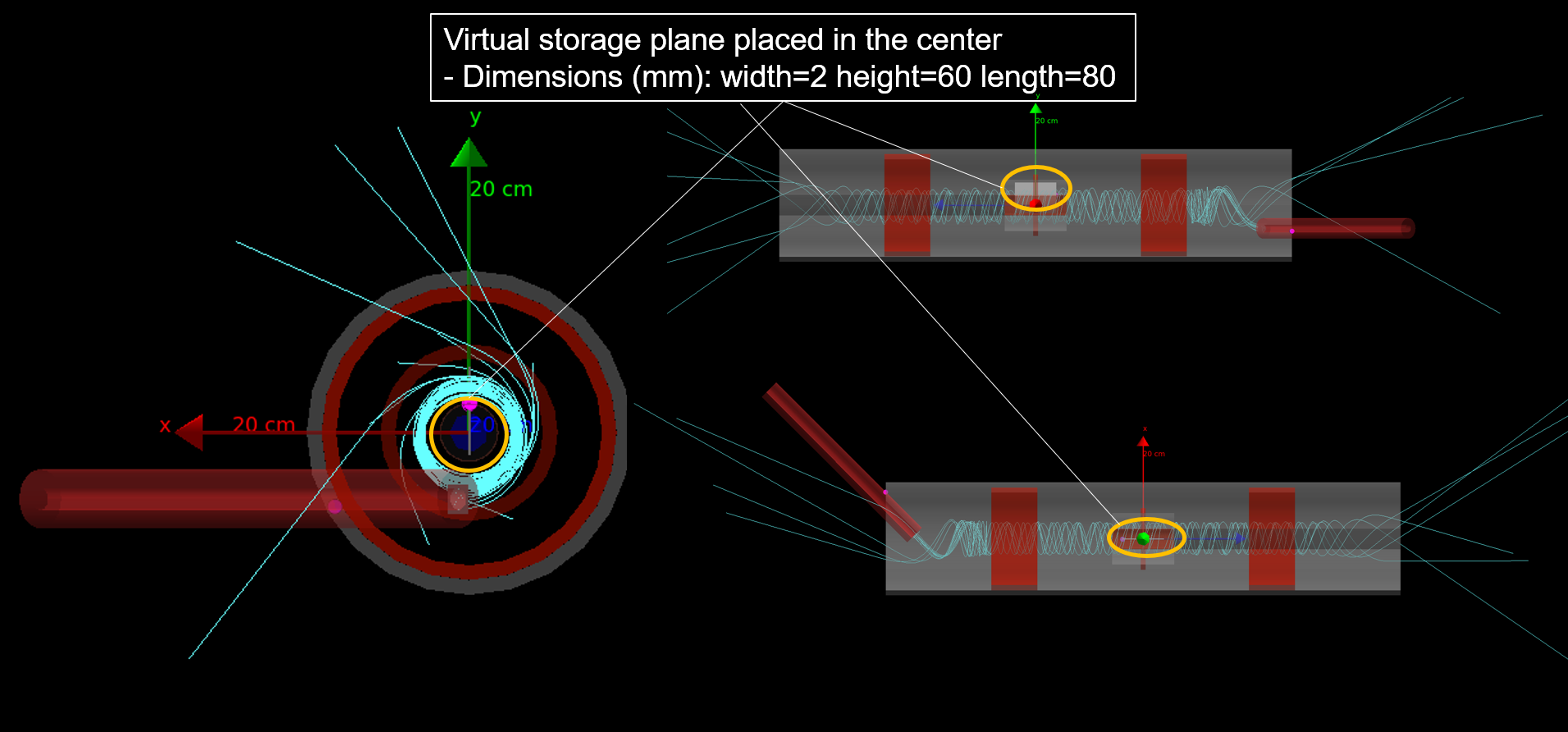}
    \caption{Visualisation of virtual storage plane implemented in G4beamline. The orange ellipse highlights the virtual storage plane in a white rectangle.}
    \label{fig:visG4BL}
\end{figure}

\begin{figure}
    \centering
    \includegraphics[width=\linewidth]{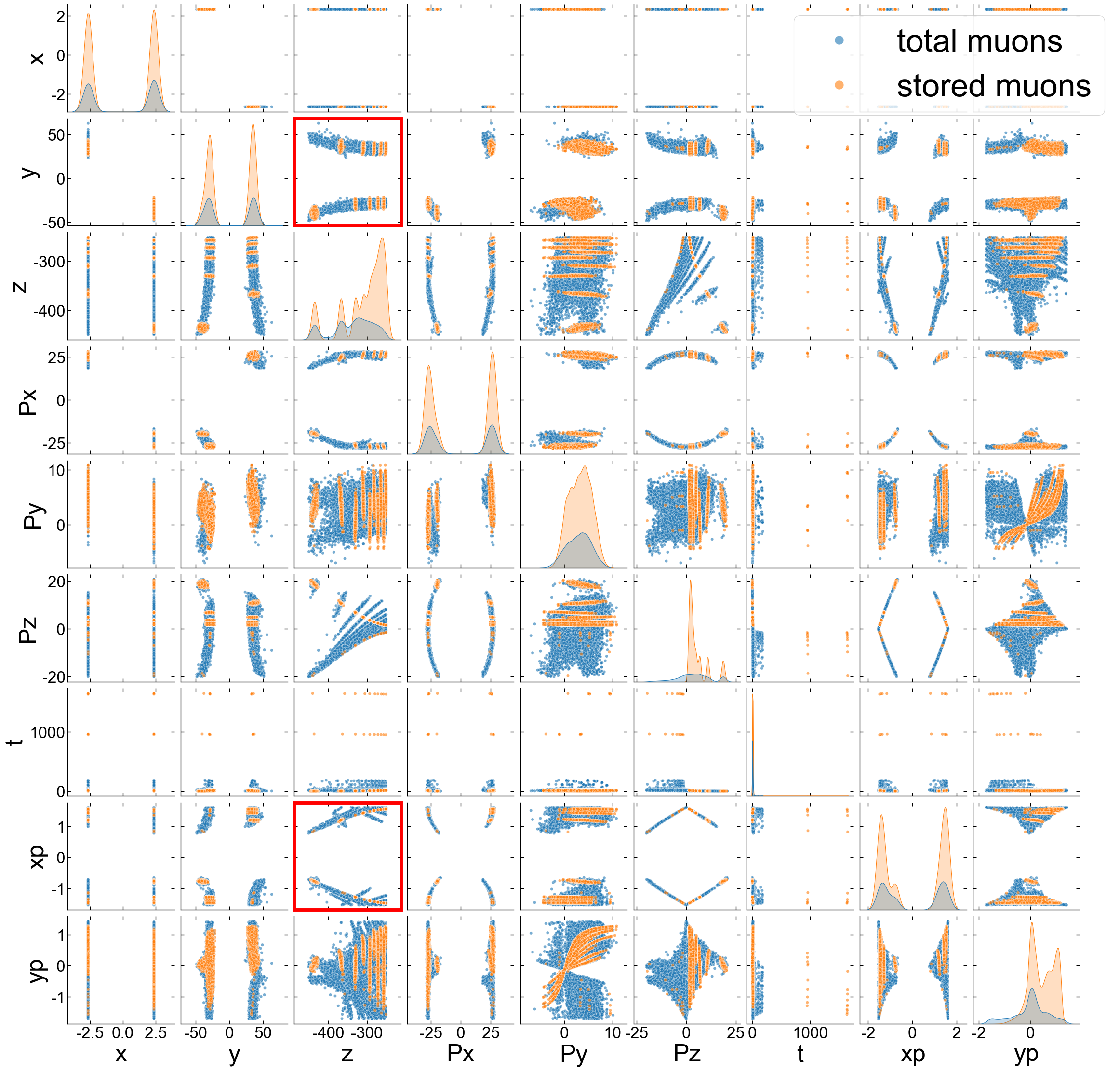}  
    \caption{Pair-plots of phase space correlations for the Aperture. Red boxes highlight key phase space correlations informing Aperture geometry.}
    \label{fig:pairplot}
\end{figure}

With these parameters implemented in G4Beamline (Fig.~\ref{fig:visG4BL}), a phase acceptance study was performed at the assumed aperture position. A virtual detector plane of dimensions $5 \times 130 \times \SI{135}{mm^3}$ was placed immediately downstream of the beam injection region. The plane was centered on the solenoid axis in the transverse plane, spanning the full extent of beam trajectories traversing this region. The resulting nine-dimensional phase space correlations at the virtual plane are presented in Fig.~\ref{fig:pairplot}.

The pair-plot analysis facilitated identification of storable muons registered on the virtual plane, which are indicated as orange points in Fig.~\ref{fig:pairplot}. To determine the key parameters governing the aperture design, particular attention was paid to the $z$--$y$ and $z$--$x'$ phase-space correlations (highlighted in red in Fig.~6), which respectively informed the aperture-opening dimensions and the angular orientations of the openings.

The simulation workflow thus established the baseline aperture design based on phase acceptance studies expected from simulations. Since $\varepsilon_a$ and $\varepsilon_r$ depend on the detector response, the workflow could be readily adapted to accommodate variations in experimental conditions. This procedure was applied to the simulated setup used in Test Beam 2024, where the setup employed a reduced beam momentum and magnetic field to accommodate prevailing logistical and engineering constraints.

\subsection{Detector construction}
\label{sec:detector_construction}

\begin{figure}[htbp]
    \centering
    \includegraphics[width=0.8\linewidth]{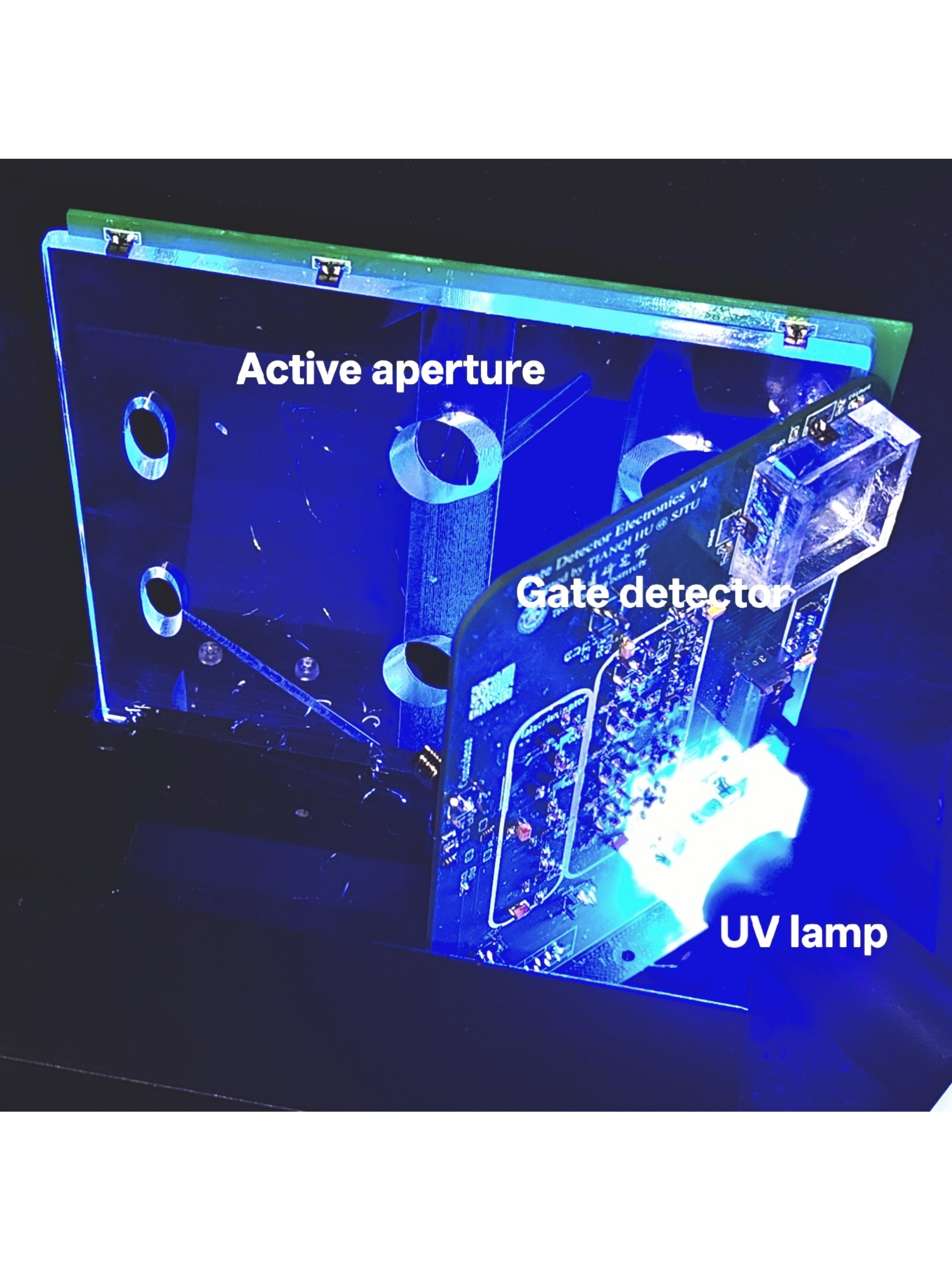}
    \caption{Picture of the assembled MTD. The Gate scintillator and its readout PCBs are visible; a UV lamp at the bottom right illuminates the Gate detector.}
    \label{fig:MTD_construct}
\end{figure}

Plastic scintillators and SiPMs were selected for the muon trigger system due to their fast response times, which were essential for meeting the experimental timing requirements. Both the Gate and Aperture were fabricated from Eljen EJ-200 plastic scintillators and were read out optically with NDL EQR20-3030 SiPMs\footnote{Novel Device Laboratory (NDL, Beijing), Devices, SiPM: \url{http://www.ndl-sipm.net/products.html\#devices}}, which feature an active area of $3.0 \times \SI{3.0}{mm^2}$. The Gate consisted of a $10 \times 10 \times \SI{0.1}{mm^3}$ thin plastic scintillator coupled to an acrylic light guide to enhance optical transmission to four SiPMs. The Aperture was a $135 \times 130 \times \SI{5}{mm^3}$ thick scintillator slab instrumented with six SiPMs, whose placements were optimized to maximize light collection from the scintillator volume.

The aperture opening geometry was derived from the acceptance map study described in Sec.~\ref{sec:sim_design} and machined using an in-house computer numerical control (CNC) module with a precision of approximately \SI{0.1}{mm}. The scintillator surfaces were coated with aluminum nano-layers to maximize scintillation light collection at the SiPMs and to prevent optical crosstalk between the Gate and Aperture. The setup includes two injection tubes to accommodate both clockwise and counter-clockwise injection directions. A custom 3D-printed holder was used to mount all detector components and readout electronics within the test beam setup, positioning the Gate and Aperture in a configuration closely matching their simulated placement. A picture of the constructed detector is shown in Fig.~\ref{fig:MTD_construct}.

\subsection{Readout electronics}

The plastic scintillators were integrated with a fast electronics readout system capable of generating an LVTTL signal in less than \SI{10}{ns}. A detailed description of the fast circuit design is provided in Ref.~\cite{Hu:2025tgk}. The readout system processed the four SiPM signals connected to the Gate, with a switch enabling alternating readout between the top and bottom channels. The Aperture signal was derived from a single readout channel with all six SiPMs connected in parallel. An onboard discriminator evaluated the anti-coincidence logic and subsequently produced the LVTTL trigger signal.

\section{Experimental setup of Test Beam 2024}
\label{sec:setup}

The muEDM Test Beam 2024 (TB24) campaign comprised a week-long run at the $\pi$E1 beamline at PSI in October 2024. The primary objectives were to characterize the MTD by measuring its detector response and to demonstrate the fast-electronics trigger capability. To accommodate evolving logistical and engineering constraints, experimental conditions were scaled down from the nominal Phase-1 design, particularly regarding the beam momentum and magnetic field configurations. Additionally, measurements were performed in air rather than vacuum, a practical compromise that nevertheless enabled our detector characterization goals.

\subsection{Beam Configurations}

\begin{figure}[htbp]
     \centering 
       \includegraphics[width=0.8\linewidth]{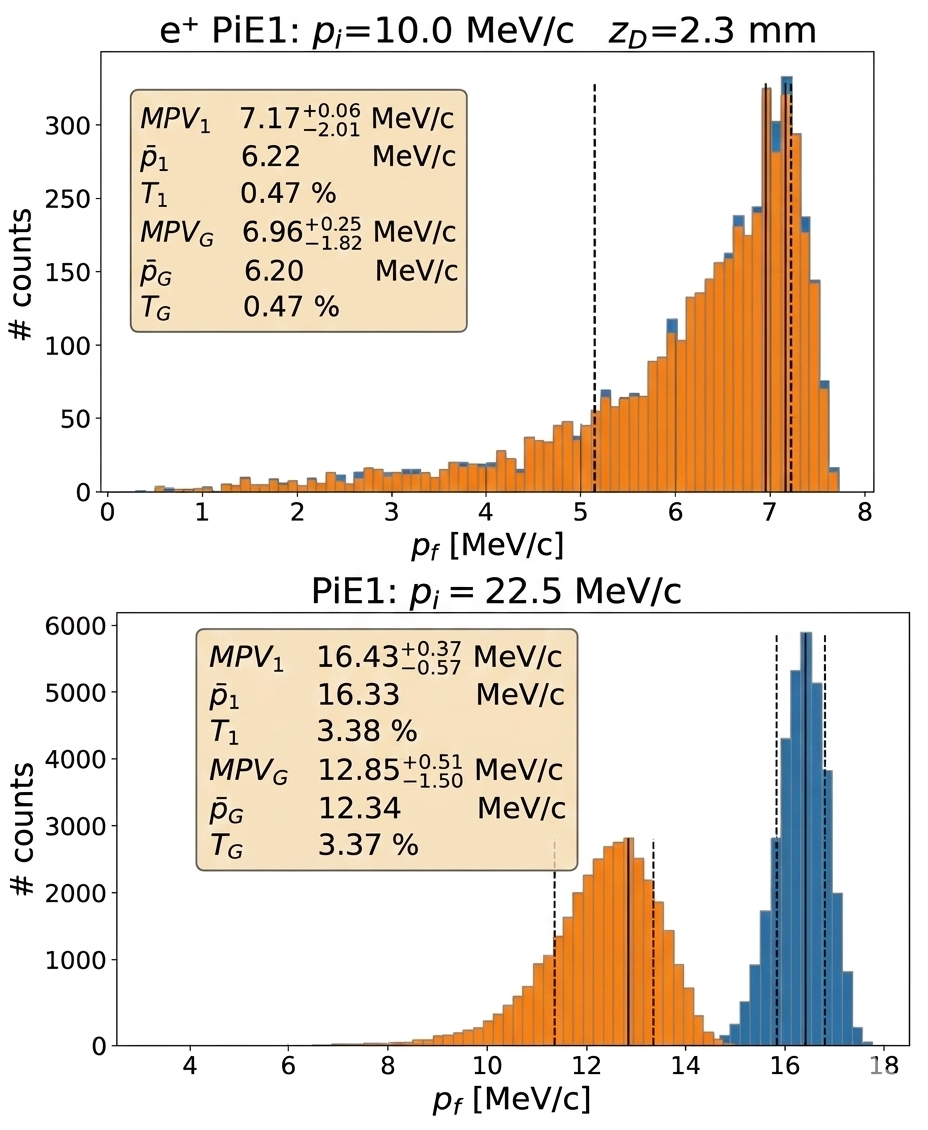}
        \caption{Simulated momentum distributions at initial momentum $p_i$, where blue and orange show before and after the Gate; the $e^+$ beam (top) traverses a copper degrader of thickness $z_D$, while no degrader is used for the $\mu^+$ beam (bottom).}
        \label{fig:tb24_beam}
\end{figure}

Two distinct beam configurations were employed to accommodate the test beam conditions in air. In the first configuration, a positron ($e^+$) beam with a momentum of \SI{7}{MeV/\textit{c}} was used; such positrons are expected to follow helical trajectories (the same as surface muons in a \SI{3}{T} magnetic field) through the aperture and reach the exit detector, thereby exercising the anti-coincidence scheme across all aperture openings. This beam condition was achieved using a combination of air, mylar, and copper degraders, with the required material thicknesses determined from simulation (Fig.~\ref{fig:tb24_beam}a). Accordingly, a \SI{2.3}{mm}-thick copper degrader was installed immediately downstream of the injection tube, just beyond the Gate, during the $e^+$ beam measurements. 

Since the trigger detector was originally optimized for muon detection, its efficiency was expected to be reduced under the $e^+$ beam configuration. To compensate for this limitation, a second beam configuration utilizing $\mu^+$ at \SI{22.5}{MeV/\textit{c}} was employed (Fig.~\ref{fig:tb24_beam}b). In this configuration, the magnetic field is off and the beam traverses only the first aperture opening. During the $\mu^+$ operation, the copper degrader was removed. The muon dataset served as the primary basis for validating the detector response in simulation.

\subsection{Experimental Setup}

\begin{figure}[htbp]
    \centering
    \includegraphics[width=0.8\linewidth]{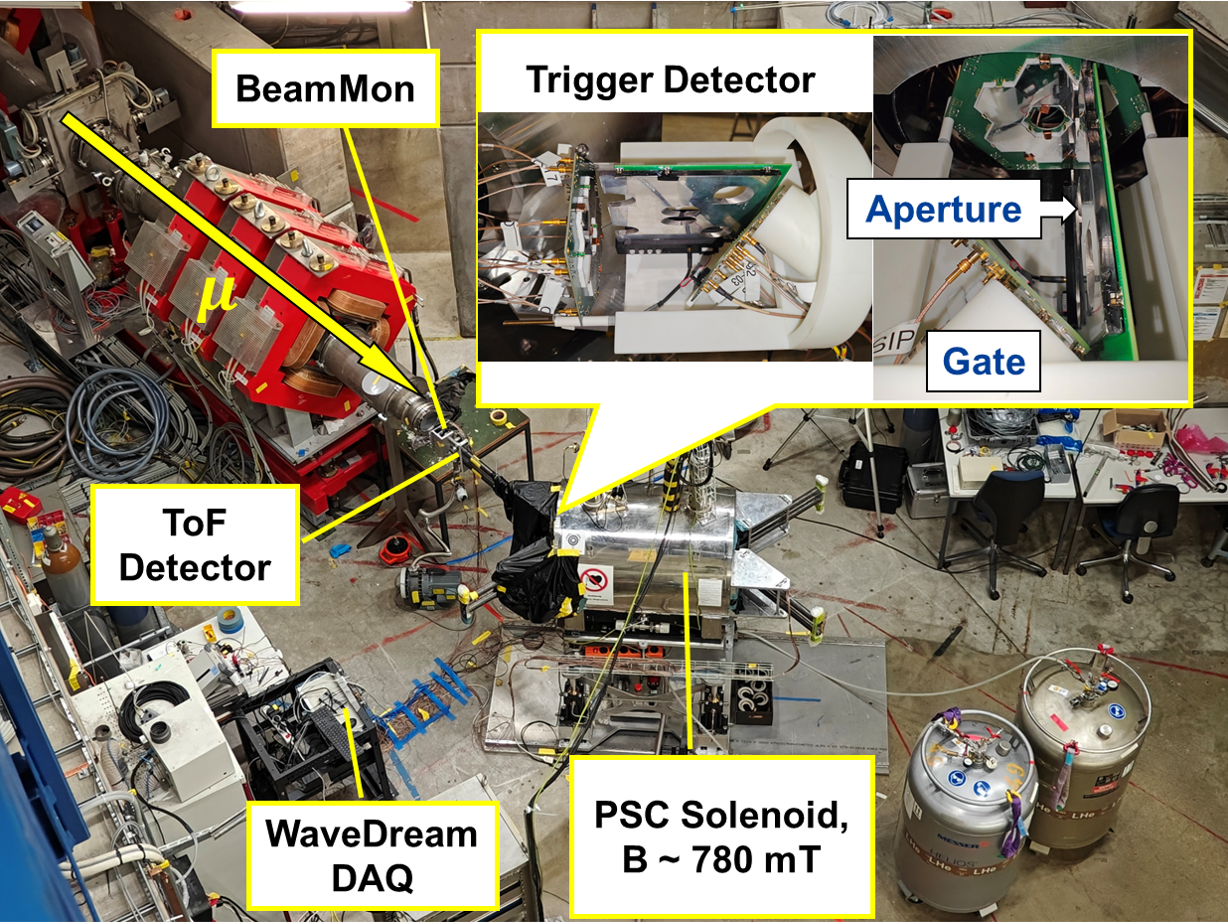} 
    \caption{Aerial view of the muEDM Test Beam 2024 experimental setup. The muon beam from $\pi$E1 enters from the top left to the PSC solenoid, where the MTD is hosted at its entrance (inset showing MTD).}
    \label{fig:tb24setup}
\end{figure}

\begin{figure}[htbp]
    \centering
    \includegraphics[width=\linewidth]{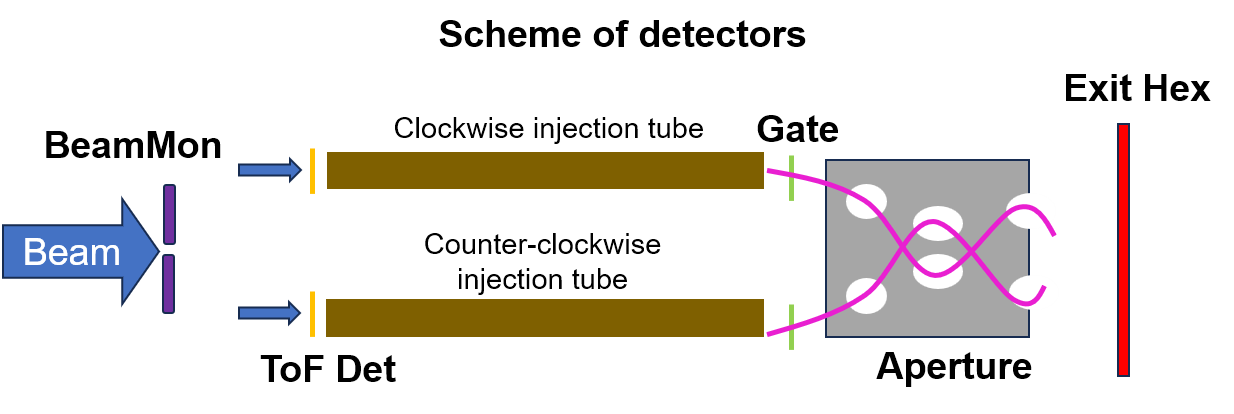}
    \caption{Schematic of detectors installed during TB24, a side view of the experimental setup, the magenta lines represent nominal $\mu^+$ trajectories.}
    \label{fig:tb24_setupScheme}
\end{figure}

The TB24 experimental setup is illustrated in Fig.~\ref{fig:tb24setup}, with the BeamMon and ToFDet positioned outside the PSC solenoid, and the MTD and ExHex installed inside (Fig.~\ref{fig:tb24_setupScheme}). Within the solenoid, the MTD comprised Gate and Aperture detectors fabricated from EJ-200 scintillator and read out using NDL EQR20-3030 SiPMs---four for the Gate, connected via a \SI{5}{mm} light guide, and six connected in parallel for the Aperture. Optical coupling for all SiPMs was achieved using BC-603 optical grease.

\subsection{DAQ}

The data acquisition (DAQ) system employed during the beam test was the WaveDream Board (WDB)~\cite{Francesconi:2023cxt}, a 16-channel waveform digitizer module developed at PSI for the MEG~II experiment. The WDB features an integrated power supply and amplification stage, enabling precise readout of fast detector signals within a compact, multi-channel architecture well suited for prototyping the MTD.

\begin{figure}[htbp]
     \centering 
       \subfigure[]{\includegraphics[width=0.54\hsize]{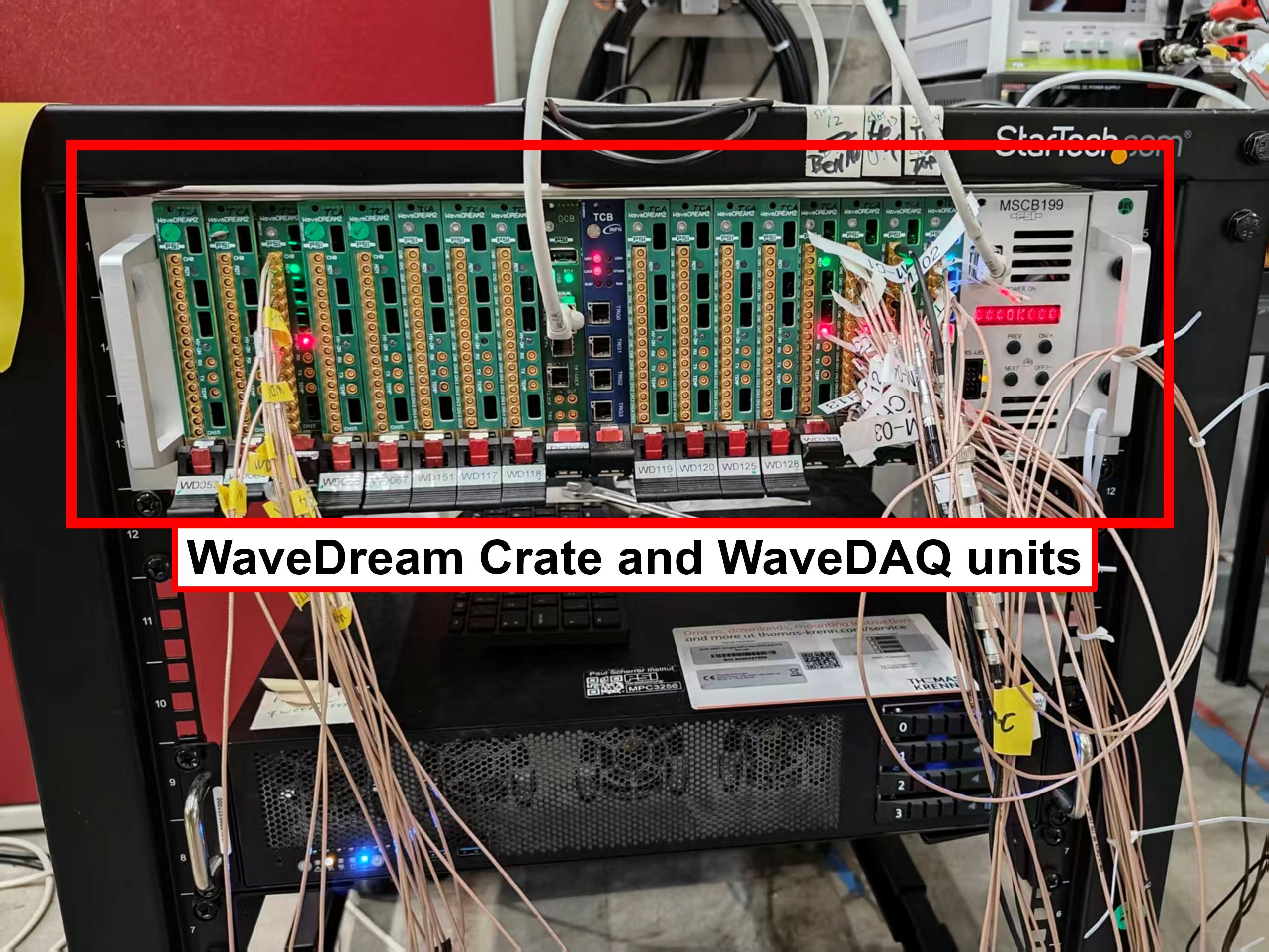}}
       \subfigure[]{\includegraphics[width=0.42\hsize]{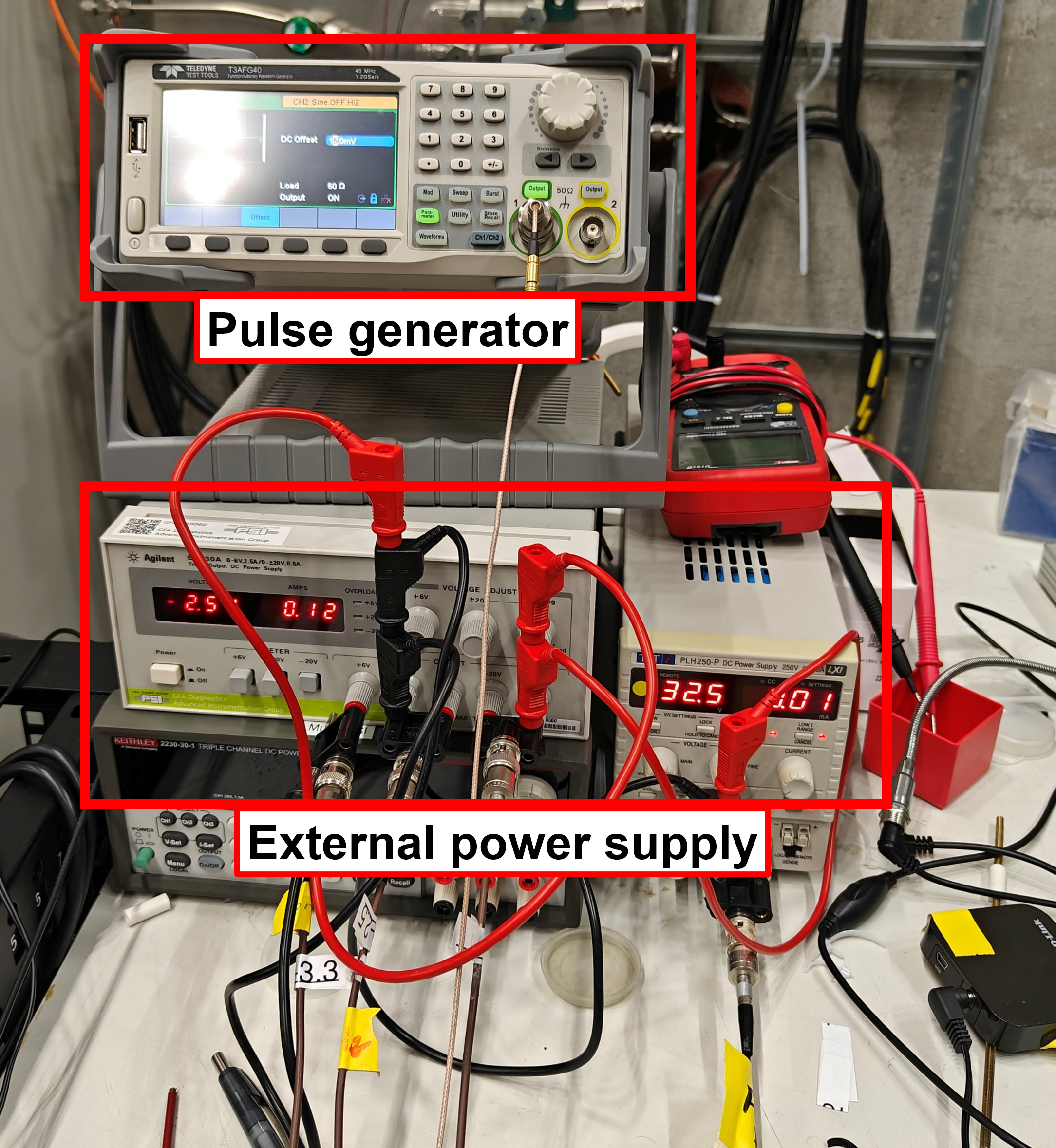}}
        \caption{a) The WaveDream Crate with three WaveDAQ online for data-taking; b) external power supply units and pulse generator for the TrigDet.}
        \label{fig:tb24_wdb}
\end{figure}

All SiPM channels were digitized using the WaveDAQ system (Fig.~\ref{fig:tb24_wdb}a). Except for the MTD channels, all detector channels were powered via the WaveDream Crate. The MTD SiPMs and preamplifier were powered by an external supply, with a pulse generator used to set the Gate's onboard TTL trigger threshold.

\begin{figure}[htbp]
    \centering
    \includegraphics[width=\linewidth]{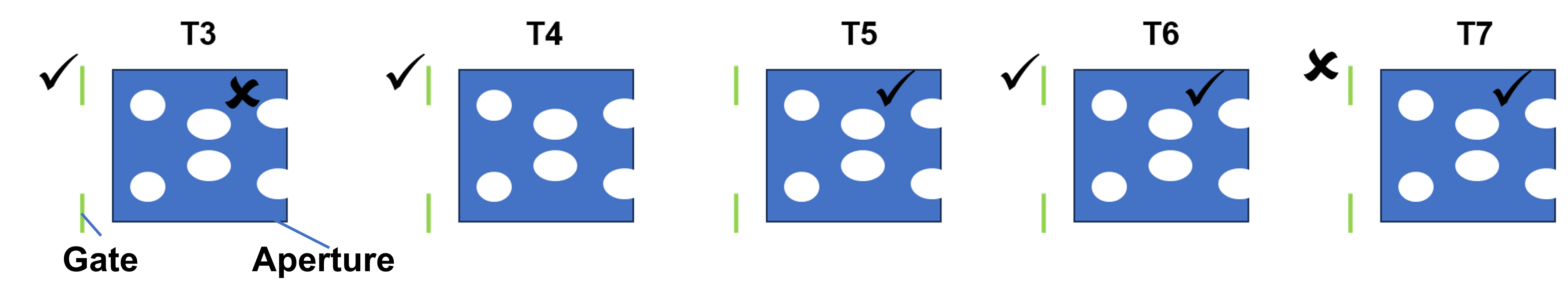}
    \caption{Scheme of each trigger pattern. The checkmark represents a registered hit, while the cross represents no hit registered.}
    \label{fig:trigPatt}
\end{figure}

Approximately $3 \times 10^6$ effective events were recorded across 140 runs (${\sim}\,1 \times 10^6~e^+$, ${\sim}\,2 \times 10^6~\mu^+$) using five trigger patterns (Fig.~\ref{fig:trigPatt}):
\begin{enumerate}
    \item \textbf{T3: LVTTL signal:} Hardware-triggered by the MTD's anti-coincidence logic.
    \item \textbf{T4: Gate self-trigger:} Gate-only detection via triple SiPM coincidence.
    \item \textbf{T5: Aperture self-trigger:} Aperture-only detection.
    \item \textbf{T6: Gate--Aperture coincidence:} Simultaneous Gate (triple SiPM) and Aperture hits.
    \item \textbf{T7: Aperture--Gate anti-coincidence:} Aperture hits with no Gate signal (triple SiPM).
\end{enumerate}

All trigger thresholds were defined within the MIDAS online trigger system, except for T3, which used a hardware discriminator threshold.

\section{Results and discussion}
\label{sec:results}

\subsection{Detector response and particle identification}
\label{sec:response}

The measured detector response during the beam test yielded important insights into the light yield and SiPM characteristics of both the Gate and Aperture subsystems. These measurements served as inputs for the simulation model, facilitating evaluation of simulation accuracy and enabled iterative refinement to enhance detector performance.

\begin{figure}[htbp]
\centering 
\subfigure[]{\includegraphics[width=.49\hsize]{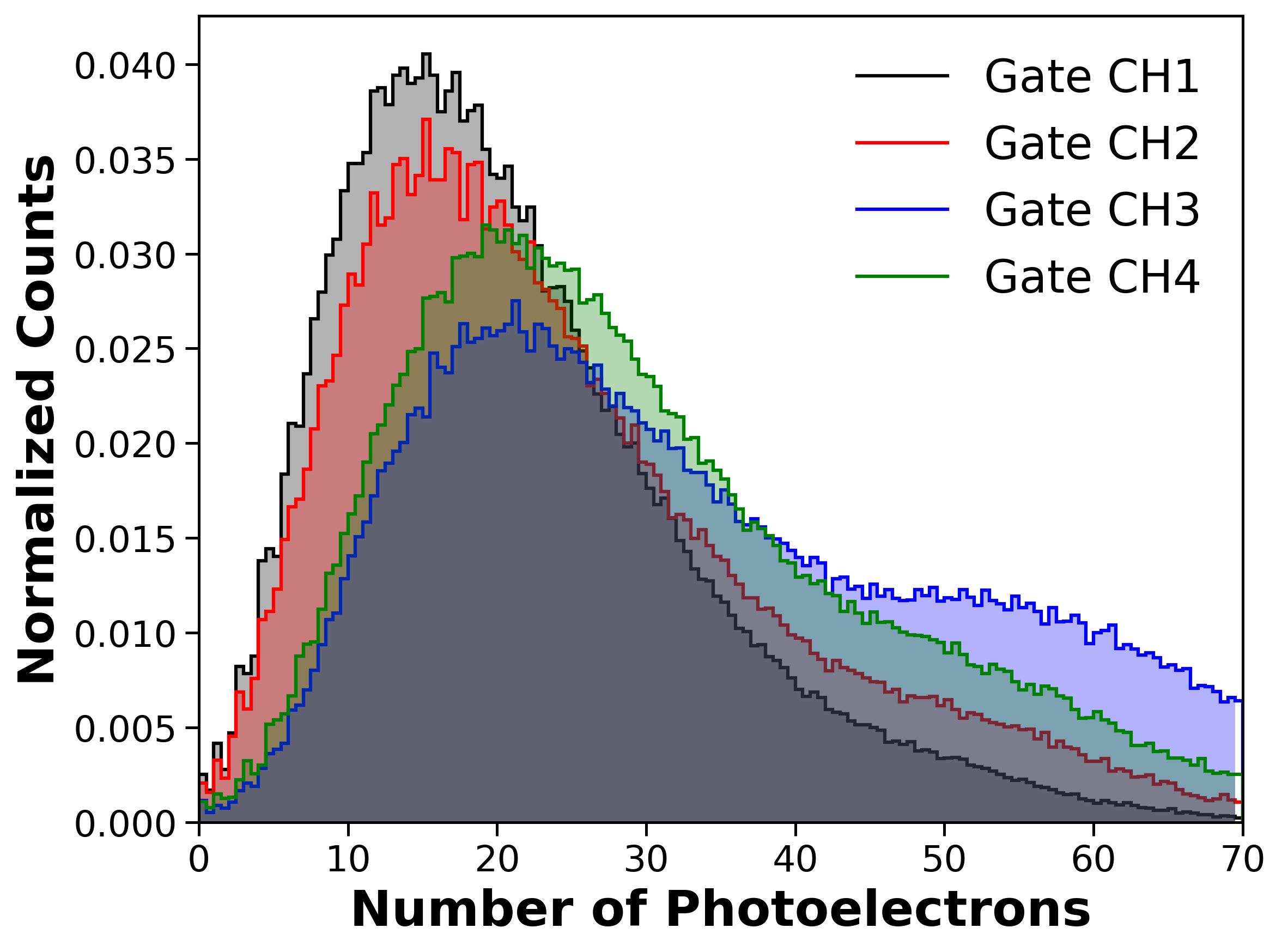}}
\subfigure[]{\includegraphics[width=.49\hsize]{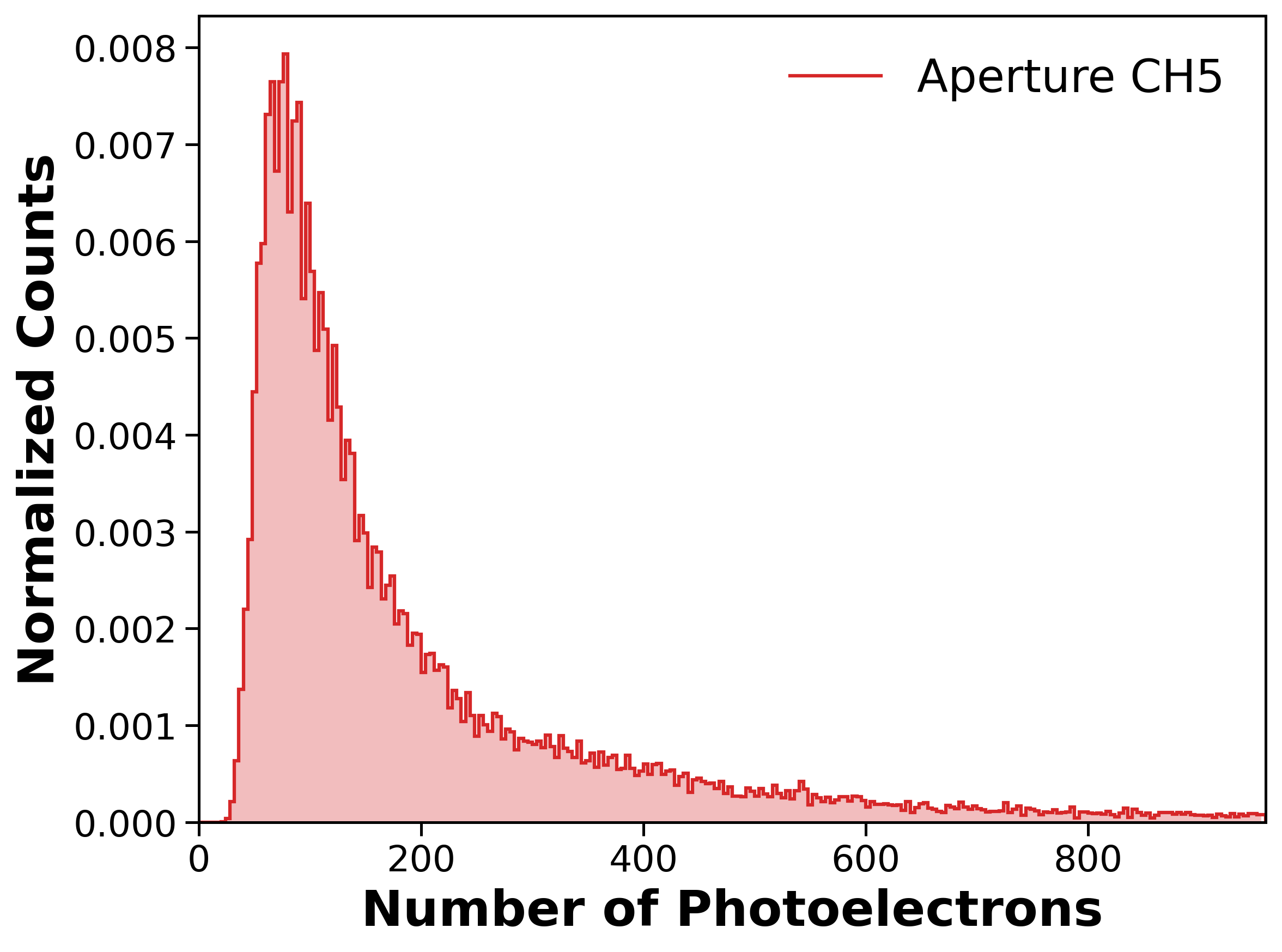}}
\caption{(a) Measured charge distributions from the four SiPM channels of the Gate; (b) measured charge distribution of the Aperture detector, from six SiPMs connected in parallel.}
\label{fig:Charge}
\end{figure}

Figure~\ref{fig:Charge} presents the measured charge distributions for each detector subsystem. These distributions are derived from the datasets exhibiting the highest expected event yield for each sub-detector component: the T4 Gate self-trigger dataset for the Gate's light yield distribution, and the T5 Aperture self-trigger dataset for the Aperture's optical response. To convert measured charge signals into photoelectron (p.e.) counts, charge calibration data for the Gate were applied. For the Aperture, the charge-to-photoelectron conversion factor was obtained from the simulation parameters detailed in Sec.~\ref{sec:simval}.

At the Gate SiPMs, the measured optical photon distribution exhibited a Landau profile, characteristic of muons traversing the thin scintillator. All four channels displayed a peak in the range of 15--20~p.e., accompanied by a long tail extending toward higher values. Notably, CH3 exhibited a broader distribution with a more prominent high-energy tail, indicative of higher-energy deposition or enhanced light collection in that channel. The observed variations in peak positions and widths among channels reflected differences in optical coupling efficiency or geometric alignment.

In contrast, the Aperture response is expected to follow a Gaussian distribution, since the \SI{22.5}{MeV/\textit{c}} muons are fully stopped within the \SI{5}{mm}-thick scintillator. However, the T5 trigger dataset was acquired with an online threshold of \SI{35}{mV}, thereby suppressing signals below this threshold. Consequently, the observed Aperture profile in Fig.~\ref{fig:Charge} appears as a truncated Gaussian, reflecting the exclusion of low-amplitude events imposed by this threshold.

\begin{figure}[htbp]
    \centering
        \subfigure[]{
        \includegraphics[width=.46\linewidth]{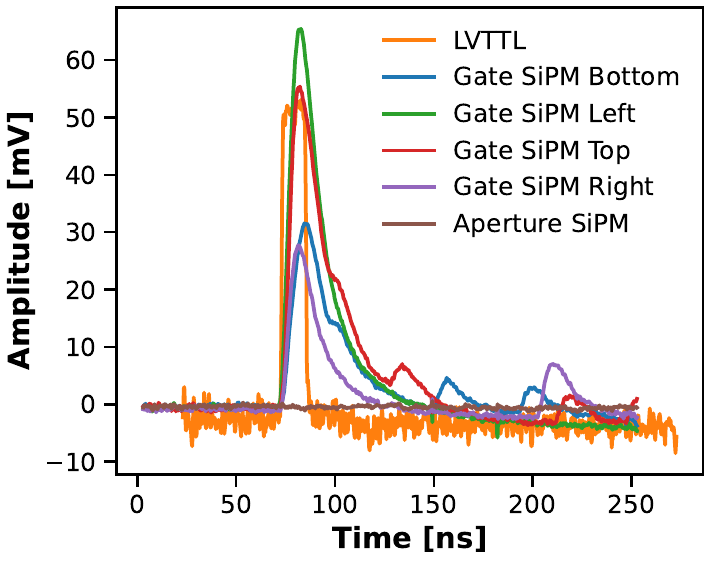}}
       \quad
    \subfigure[]{
        \includegraphics[width=.46\linewidth]{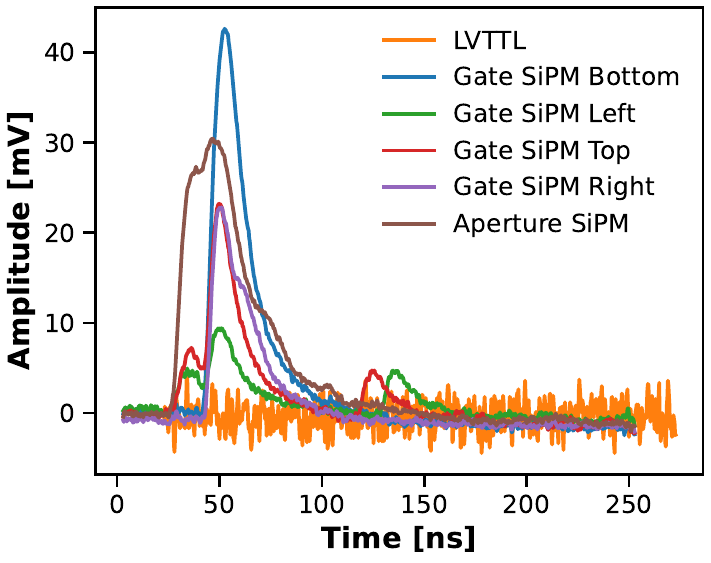}}
    \caption{a) Waveform example of an accepted event showing signals from all four Gate SiPMs (orange) and no signal from the Aperture SiPM, along with the generated TTL signal; b) waveform example of a rejected event.}
    \label{fig:typicalWaveform}
\end{figure}

Figure~\ref{fig:typicalWaveform} illustrates representative waveforms for accepted events (Gate anti-coincidence) and rejected events (Gate--Aperture coincidence).

\subsection{Event Topology}
\label{sec:topology}

\begin{figure}[htbp]
    \centering
    \includegraphics[width=0.7\linewidth]{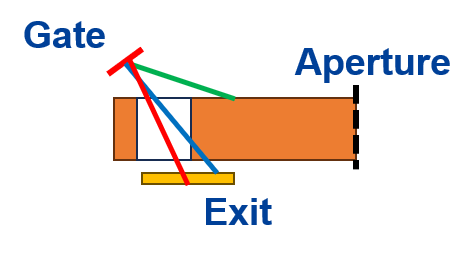}
    \caption{A top-view schematic illustration of MTD-TB24 shows the event topology configuration for the Gate self-trigger T4 dataset, where colored trajectories distinguish different event categories. The red lines represent accepted events that pass through the Gate without interacting with other detectors, while green lines show rejected events where particles hit both the Gate and Aperture. Blue lines indicate additional rejected events that register hits in the Gate, Aperture, and Exit detectors.}
    \label{fig:evtopo_scheme}
\end{figure}

Utilizing the detector response information, the event topology was analyzed to quantify the MTD's selective triggering capability. Following the methodology established for the proof-of-concept design~\cite{Hu:2025egg}, an event classification chain categorized events based on the trigger patterns of the participating sub-detectors. The resulting expected event topologies are illustrated in Fig.~\ref{fig:evtopo_scheme}, with the corresponding event rates summarized in Table~\ref{tab:topology}.

As presented in Table~\ref{tab:topology}, approximately 2\% of events registered in the Aperture, while approximately three-quarters of all events satisfied the anti-coincidence condition. This outcome is consistent with design expectations: the aperture geometry was initially optimized for a lower-momentum $\mu^+$ beam ($p = \SI{7}{MeV/\textit{c}}$) exhibiting broader angular scattering (see Sec.~\ref{sec:design}). By contrast, the TB24 beam configuration employed a higher-momentum $\mu^+$ beam ($p = \SI{22.5}{MeV/\textit{c}}$), which underwent substantially less multiple scattering at the Gate. Consequently, the beam retained a higher degree of collimation relative to the aperture dimensions, resulting in fewer muons intersecting or stopping within the Aperture than originally anticipated.

\begin{table}[htbp]
\centering
\caption{Two-way contingency table of relative event rates calculated from $\mu^+$ runs in the T4 Gate self-trigger dataset. The ``No Aperture'' and ``No Exit'' labels denote events with no registered hits in the respective detectors.}
\label{tab:topology}
\begin{tabular}{@{}lccc@{}}
\toprule
 & Exit (\%) & No Exit (\%) & Total (\%) \\
\midrule
Aperture    & 1.69  & 2.20  & 3.89  \\
No Aperture  & 74.90 & 21.19 & 96.09 \\
\midrule
Total       & 76.59 & 23.39 & 100   \\
\bottomrule
\end{tabular}
\end{table}

\subsection{Simulation verification}
\label{sec:simval}

Building upon the proof-of-concept validation, the measurements were verified through a Geant4-based simulation. Agreement between measurement and simulation confirmed the MTD's performance and further substantiated the reliability of the detector development methodology.

\begin{figure}[htbp]
    \centering 
    \subfigure[]{
        \includegraphics[width=.49\linewidth]{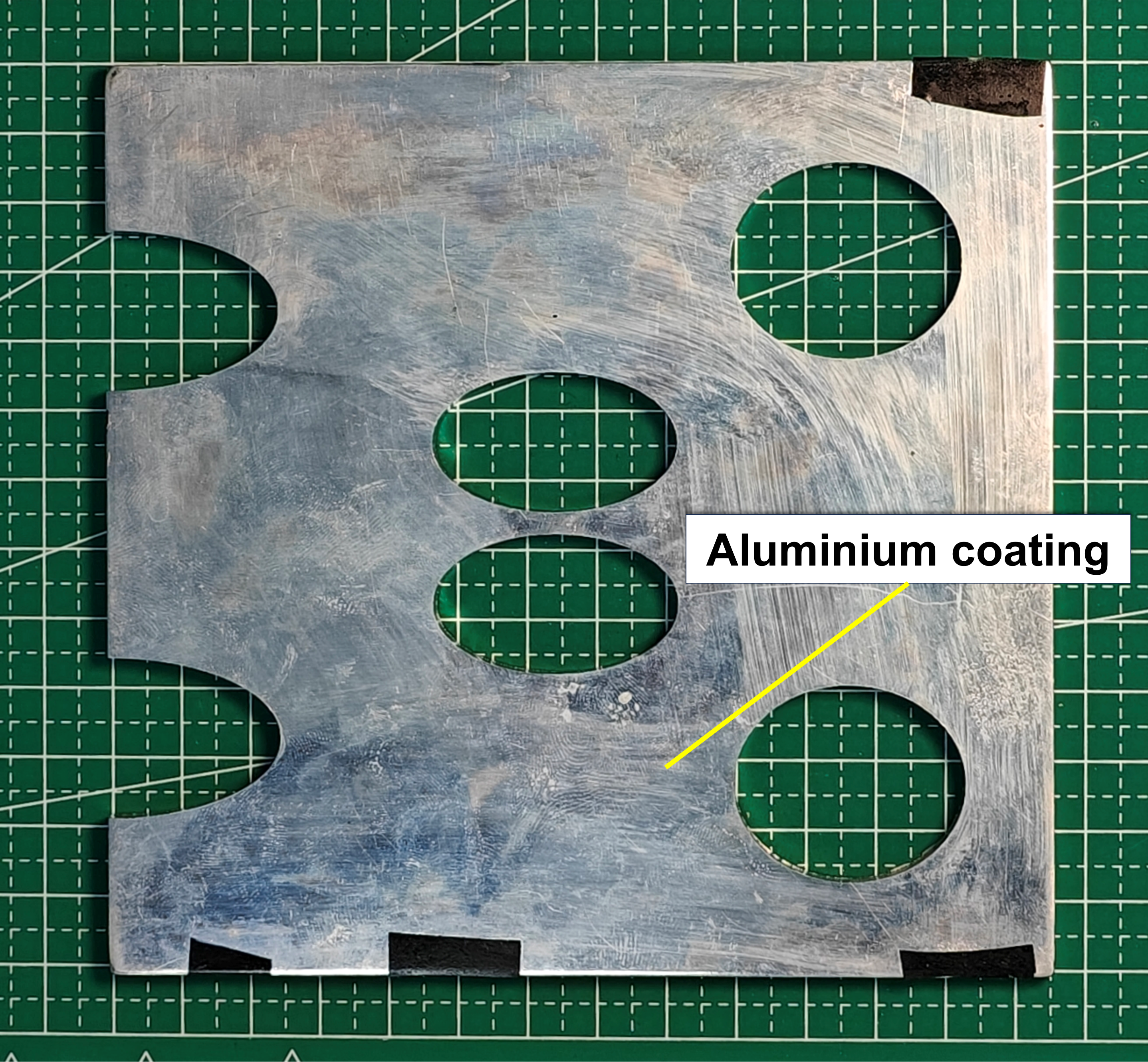}}
    \subfigure[]{
        \includegraphics[width=.45\linewidth]{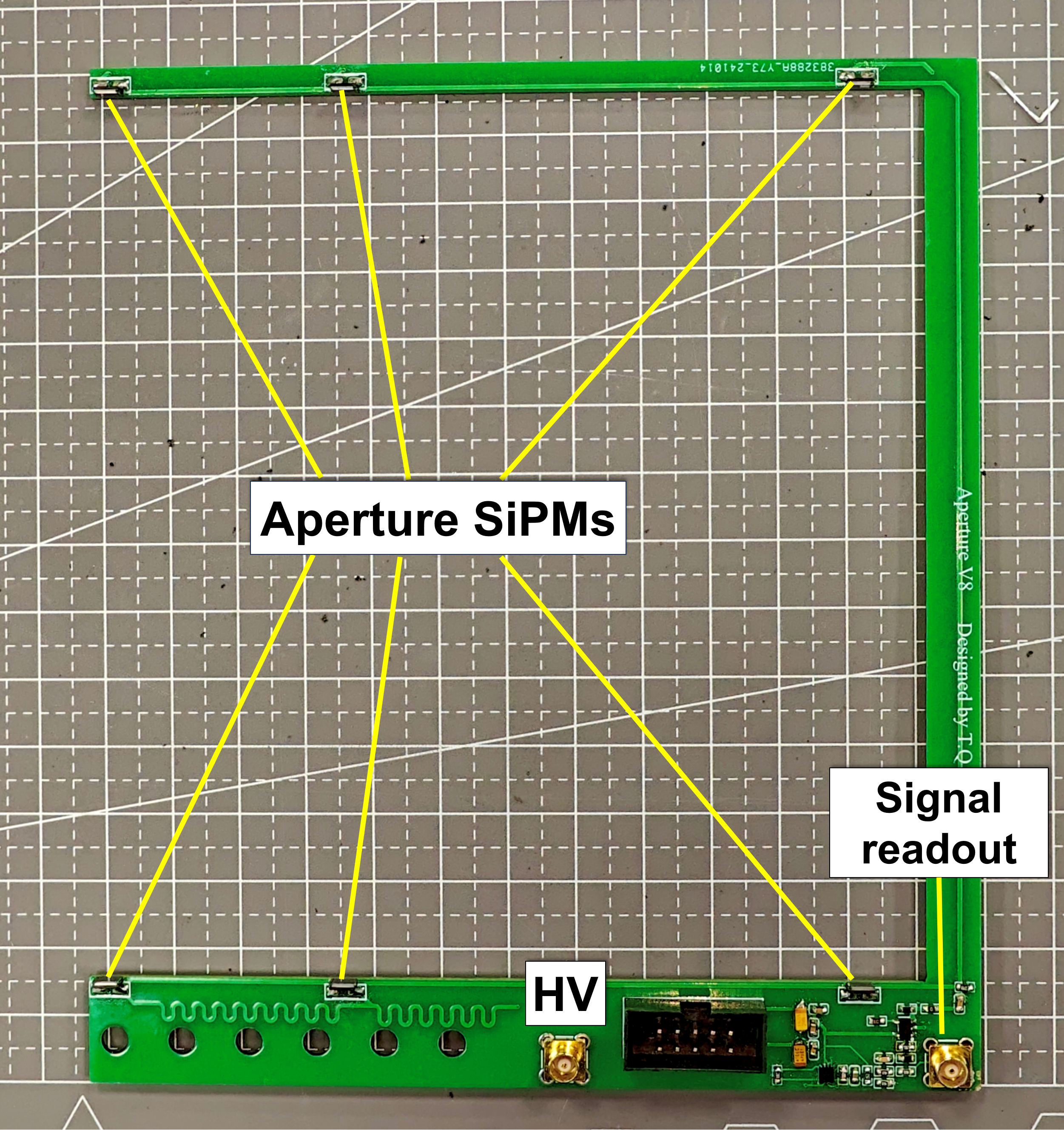}}
    \caption{a) The Aperture scintillator sputtered with an aluminium coating; b) the Aperture printed circuit board (PCB) illustrating the relative positioning of the SiPM components.}
    \label{fig:AAobj}
\end{figure}

The detector configuration in the simulation faithfully reproduced the TB24 experimental setup. For the Aperture subsystem, the SiPMs were modeled using a computer-aided design (CAD) representation (Fig.~\ref{fig:AAobj}a), enabling accurate alignment with the printed circuit board (PCB) layout (Fig.~\ref{fig:AAobj}b). To accurately replicate the detector's electrical configuration, the six Aperture SiPMs were modeled as a single photosensitive element, effectively simulating their parallel connection and combined signal output. Similarly, CAD models of the Gate SiPMs were incorporated into the simulation environment.

\begin{figure}[htbp]
    \centering
    \includegraphics[width=\linewidth]{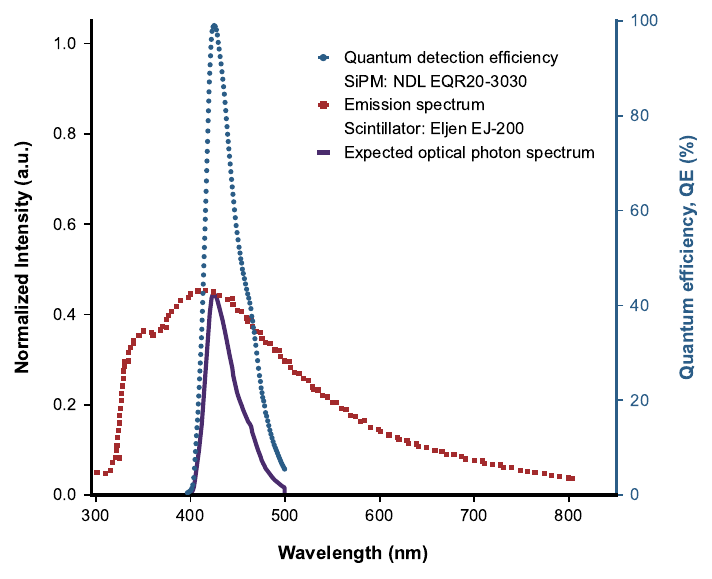}
    \caption{Wavelength distribution of the scintillation photons from Eljen plastic scintillator and the corresponding quantum detection efficiency of the NDL SiPM, alongside the expected spectrum of their optically coupled system.}
    \label{fig:tb24simval_wavelength}
\end{figure}

The detector's optical response was validated by reproducing the measured photon distributions. Manufacturer-provided data-sheet specifications for the SiPMs and plastic scintillators were integrated into the optical simulation. On the basis of the simulation validation performed for the proof-of-concept MTD, the following essential optical parameters were identified:

\begin{itemize}
    \item Scintillation yield: 10{,}000 photons/MeV
    \item Emission spectrum and photon detection efficiency (PDE) curve (see Fig.~\ref{fig:tb24simval_wavelength})
    \item Optical model: Lookup Table (LUT)
    \item Surface type: \texttt{dielectric\_metal}
    \item Surface finish: \texttt{polished}
\end{itemize}

The Gate and Aperture detectors share identical optical surface characteristics, with a critical distinction at the interface between the plastic scintillator and the surrounding medium (G4~AIR). For the Gate, no reflective layer is applied at this interface. Conversely, the Aperture features an aluminum nano-coating on its scintillator surface, which facilitates near-total internal reflection of scintillation photons. To accurately model this configuration, the boundary between the Aperture and the surrounding volume is defined as a \texttt{dielectric\_metal} interface with a \texttt{polished} finish, functioning as an ideal mirror that reflects all incident photons. This ensures confinement of optical photons within the Aperture volume and prevents crosstalk with the Gate.

\begin{figure}[htbp]
    \centering 
    \subfigure[]{
        \includegraphics[width=.48\linewidth]{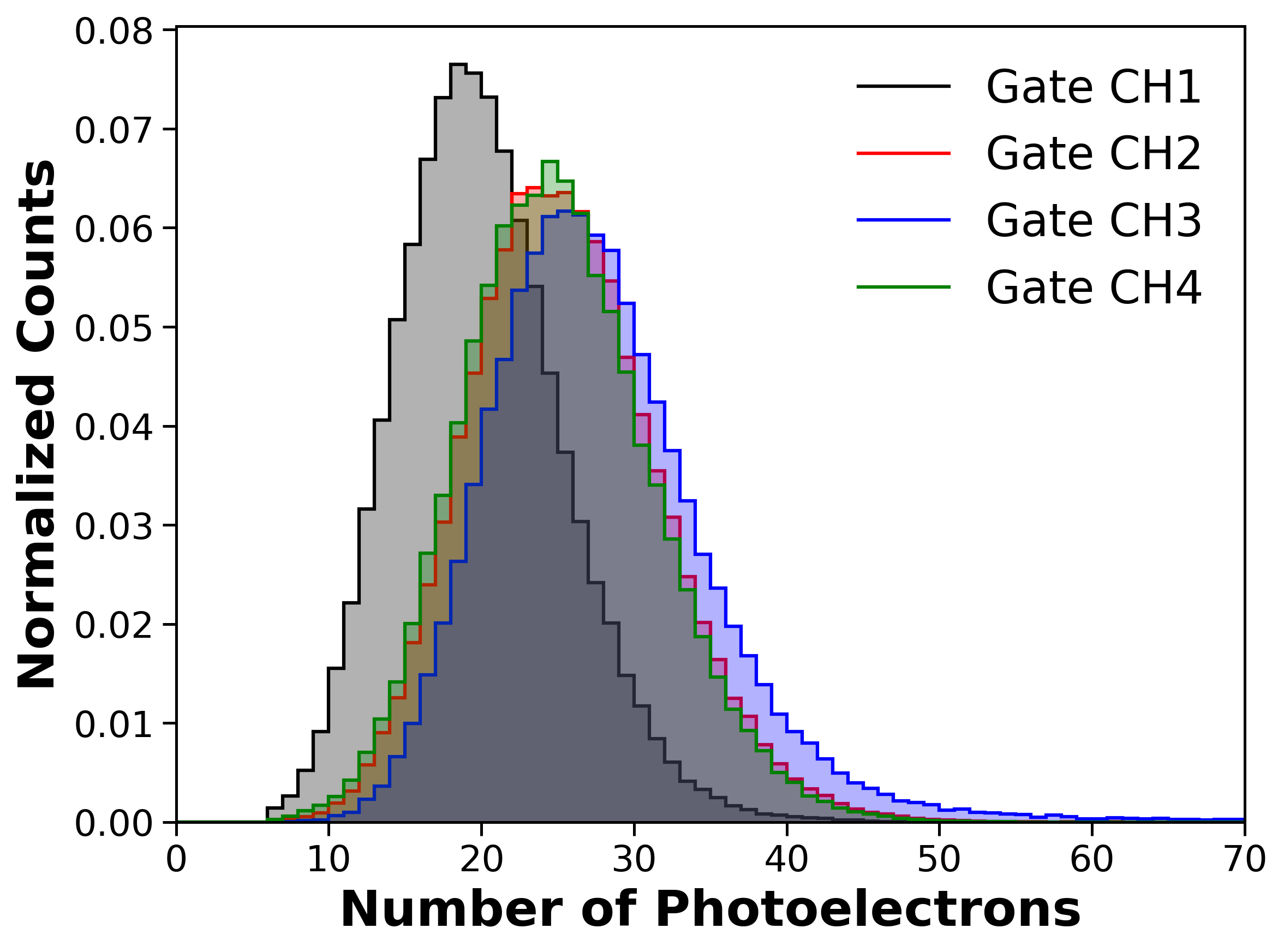}}
    \subfigure[]{
        \includegraphics[width=.48\linewidth]{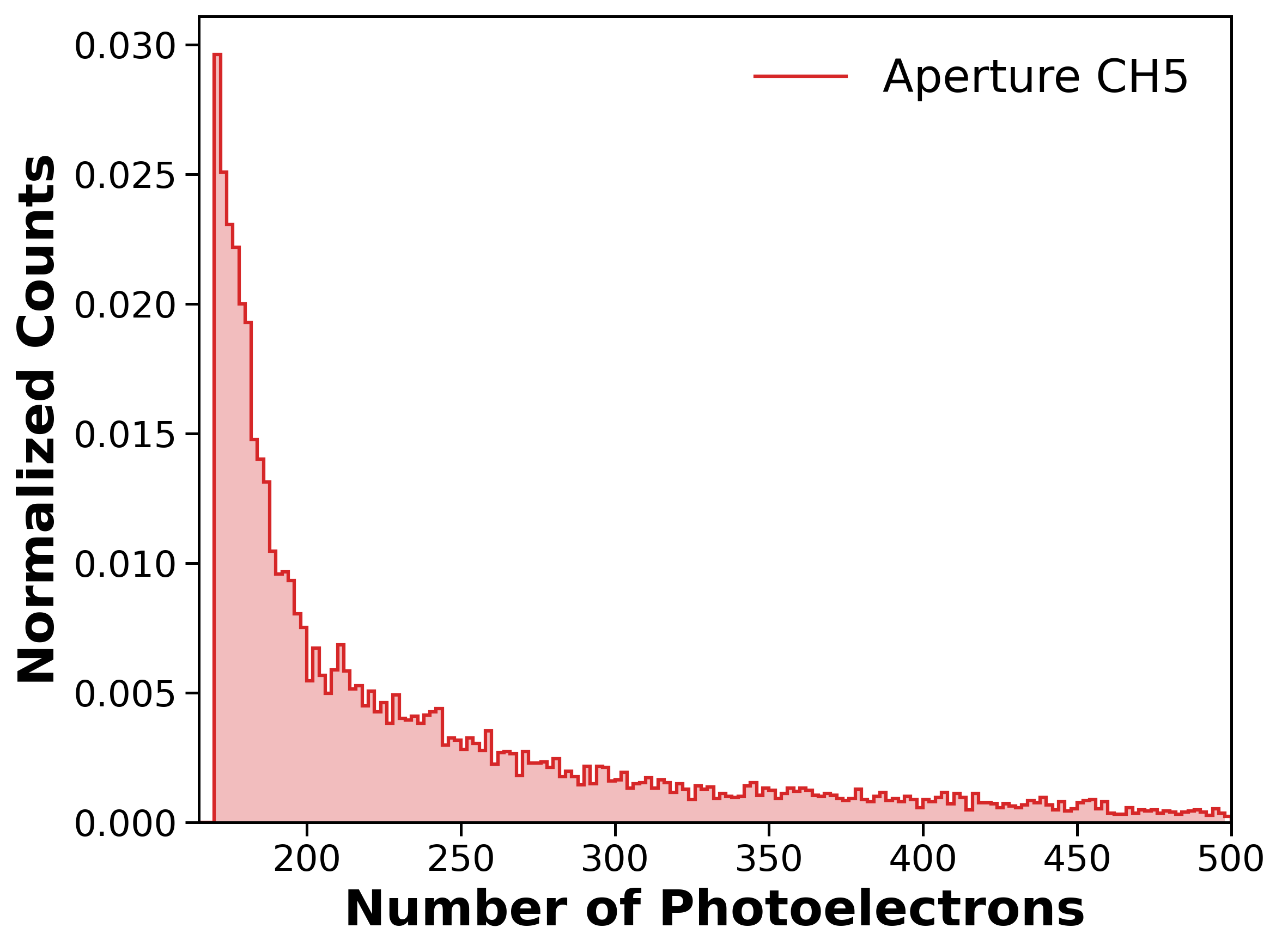}}
    \caption{Simulated optical photon distribution detected at the (a) Gate and (b) Aperture, matching the measured distributions in Fig.~\ref{fig:Charge}.}
    \label{fig:optdist_sim}
\end{figure}

All simulation parameters were meticulously calibrated to reproduce the measured distributions presented in Fig.~\ref{fig:Charge}. The spectral characteristics of detected optical photons were systematically matched by adjusting the quantum efficiency of the SiPMs and the scintillation yield of the plastic scintillator. Channel-to-channel variations in the average number of detected optical photons at the Gate SiPMs were precisely reproduced through modifications to the transmittance of the respective SiPM--scintillator interfaces. For the Aperture's optical photon distribution, a threshold cut scan was performed to align with the online trigger threshold. Fig.~\ref{fig:optdist_sim} presents the resulting simulated distributions for the Gate and Aperture, demonstrating agreement with the measured distributions in Fig.~\ref{fig:Charge}.

\begin{figure}
    \centering
    \includegraphics[width=\linewidth]{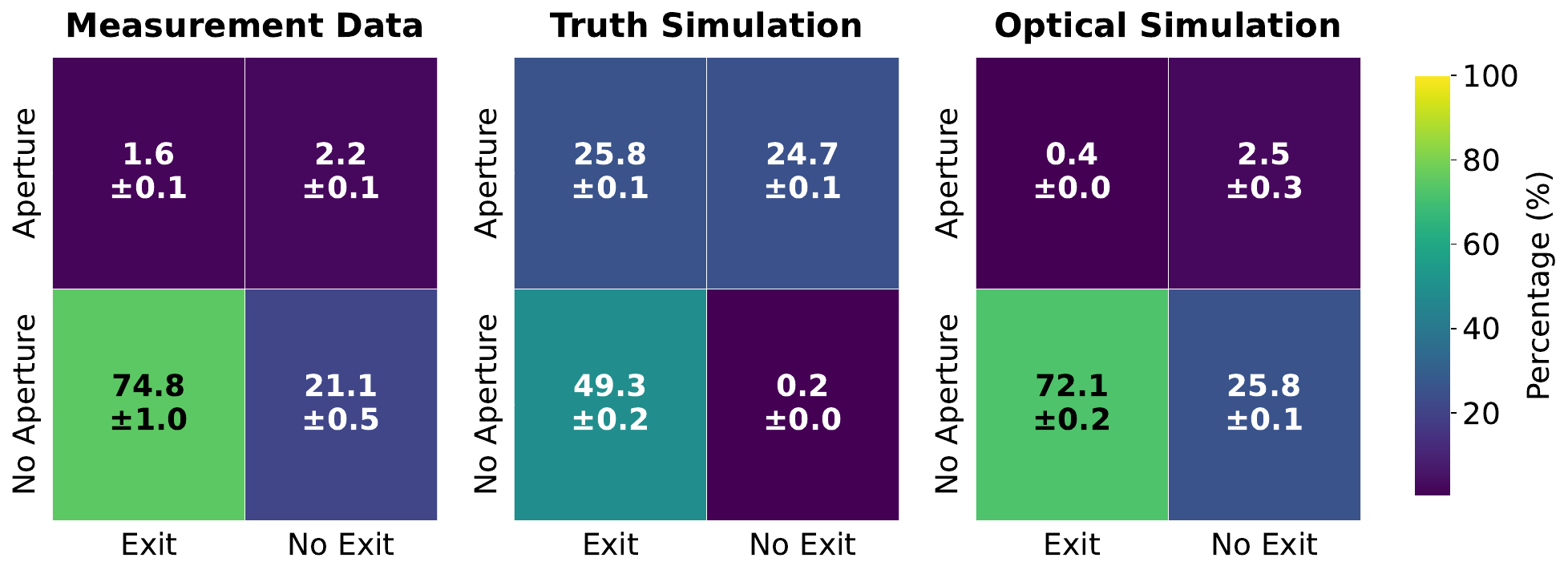}
    \caption{Heatmaps representing the comparison of event fractions between measurement, truth simulation, and optical simulation.}
    \label{fig:heatmap}
\end{figure}

To further validate the understanding of the detector response, an event topology analysis was performed for direct comparison with the measured event fractions. Figure~\ref{fig:heatmap} presents a comparison of event fractions obtained from measurements, truth-level simulation, and optical simulation. In this context, the truth-level simulation registers a hit for any nonzero energy deposit in the detectors, whereas the optical simulation additionally incorporates detector-specific optical processes---including photon transport and SiPM response---to model realistic signal generation.

By optimizing the optical photon-number thresholds applied to the simulation output, the agreement with measurements is markedly improved. This improvement is particularly pronounced for event topologies associated with anti-coincidence events and for those registering no signal in either the Aperture or Exit detector, as shown in the bottom-left and bottom-right cells of Fig.~\ref{fig:heatmap}. Whereas the truth-level simulation underestimates the measured anti-coincidence event rate by at least 20\%, the optical simulation reduces this discrepancy to within 3\%. Furthermore, the truth-level simulation entirely fails to reproduce the ``No Aperture \& No Exit'' event topology, predicting a rate of nearly 0\%, whereas the optical simulation estimates approximately 4\%, which is substantially closer to the observed values.

\section{Conclusion}
\label{sec:conclusion}
We have developed, constructed, and characterized a prototype Muon Trigger Detector for the PSI muEDM experiment. The detector, comprising a thin plastic scintillator Gate and a thick, CNC-machined plastic scintillator Active Aperture---both read out by SiPMs---was designed to provide a fast, selective trigger for storable muons via anti-coincidence logic.

The detector design was rigorously informed by a comprehensive Geant4-based simulation workflow encompassing detailed modeling of optical photon transport and SiPM response. The prototype was successfully tested with a \SI{22.5}{MeV/\textit{c}} $\mu^+$ beam during the PSI Test Beam 2024 campaign. The analysis, which focused on event topology classification, optical photon distributions, and energy deposition profiles, yielded the following key findings:

\begin{itemize}
    \item The measured detector response was successfully reproduced by the detailed optical simulation. After applying optical photon thresholds and DAQ-equivalent time windows, the simulated anti-coincidence event rates agreed with measured values to within 3\%. This represents a marked improvement over truth-level simulations, which overestimated these rates by more than 20\% due to sub-threshold energy deposits that do not produce a measurable signal.

    \item The optical simulation, incorporating reflective coatings and SiPM response modeling, successfully reproduced event topologies that the truth-level simulation failed to capture. Most notably, it modeled the ``No Aperture \& No Exit'' topology at a rate of ${\sim}4\%$, consistent with the measurements, whereas the truth-level simulation predicted a rate of nearly 0\%.

    \item Under the scaled-down TB24 conditions, the system demonstrated the operational principle of the anti-coincidence trigger. The measured event rates and topologies were consistent with a rejection efficiency well above the critical 98\% threshold required for the final experiment, thereby validating the core design concept.
\end{itemize}

These results confirm a high degree of reproducibility of the MTD response in simulation, demonstrating a robust understanding of its performance and validating the methodologies established through prior prototype iterations. The development approach described herein is inherently flexible and can be adapted to accommodate variations in experimental conditions---a crucial attribute for future iterations.

The successful beam test and subsequent simulation validation demonstrate that the MTD prototype is ready for the next stage of development. The insights gained---particularly regarding the critical importance of modeling the full optical response---will directly inform the final design and commissioning of the MTD for the full muEDM Phase-1 experiment, in which it will play a central role in the search for the muon electric dipole moment.

\section*{Acknowledgments}
The collaboration is grateful for the excellent technical support and advice from F. Barchetti, M. Gantert, U. Greuter, A. Hofer, L. Kuenzi, M. Meier, and R. Senn of LTP; T. Rauber and P. Simon of CAS for supporting the test beam; and T. Höwler and his colleagues from the PSI survey group for aligning the solenoid. We thank C. Klauser from LIN for supporting us in the coating of the scintillators with aluminum. We also acknowledge the great help of A. Knecht and A. Antognini before and during our test beam times on the piE1 beam line. Finally, we express our special thanks to all colleagues from the UCN group at PSI for generously lending us electronic devices. The project is supported by the Science Foundation of China under Grant 12050410233 and the China Scholarship Council No. 202206230093. This work is also partially funded by the Swiss National Science Foundation grants No. 204118, 220487, PP00P21\_76884, TMCG-2\_213690, and receives funding from the Swiss State Secretariat for Education, Research, and Innovation (SERI) under grant No. NoMB22.00040.

\bibliography{testbeam2024}
\end{document}